\documentclass[]{eptcs}
\providecommand{\event}{G.Ciobanu, M.Koutny (Eds.): Membrane Computing and\\
Biologically Inspired Process Calculi (MeCBIC 2010)
EPTCS Preliminary Proceedings, 2010.}
\providecommand{\firstpage}{1}

\usepackage{epsfig}
\usepackage{amsfonts}
\usepackage{amssymb}
\usepackage[leqno]{amsmath}
\usepackage{xcolor}
\usepackage{subfigure}

\newcommand{\FORGET}[1]{}

\newtheorem{theorem}{Theorem}[section]

\newtheorem{remark}[theorem]{Remark}

\def\bbbr{{\rm I\!R}}

\newcommand{\CalculusShortName}{\textsc{CWC}}%{\textsc{CWM}}

\newcommand{\qqop}[1]{\mathrel{\makebox[2em]{$#1$}}}

\newcommand{\agr}{\quad\big|\quad}

\newcommand{\mydots}{\cdot\cdot\cdot}

\newcommand{\AT}{\mathcal{A}}
\newcommand{\LT}{\mathcal{L}}
\newcommand{\AS}{\overline{\mathcal{A}}}
\newcommand{\TT}{\mathcal{T}}
\newcommand{\TS}{\overline{\mathcal{T}}}

\newcommand{\Pat}{p}
\newcommand{\LeftPP}{\mathcal{P}}
\newcommand{\LeftPat}{P}

\newcommand{\OT}{\mathcal{O}}

\newcommand{\RightPP}{\mathcal{O}}
\newcommand{\RightPat}{O}

\newcommand{\TOP}{\top}
\newcommand{\LAB}{\textsc{Lab}}

\newcommand{\BR}{\textsc{BR}}

\newcommand{\VarOf}{\textit{Var}}
\newcommand{\LengthOf}[1]{|#1|}

\newcommand{\CC}{\mathcal{C}}

\newcommand{\QQ}{\mathcal{Q}}
\newcommand{\RR}{\mathcal{S}}
\newcommand{\DD}{\mathcal{D}}
\newcommand{\BB}{\mathcal{B}}
\newcommand{\Ode}{\mathcal{E}}

\newcommand{\NN}{\mathcal{N}}
\newcommand{\OO}{\textsc{Occ}}

\newcommand{\ltrans}[1]{\xrightarrow{#1}}

\makeatletter
\def\mapstofill@{%
\arrowfill@{\mapstochar\relbar}\relbar\rightarrow}
\newcommand*\xmapsto[2][]{%
\ext@arrow 0000\mapstofill@{#1}{#2}}
\makeatother

\newcommand{\srewrites}[1]{\stackrel{#1}{\longmapsto}}
\newcommand{\xsrewrites}[1]{\xmapsto{#1}}

\newcommand{\into}{\ensuremath{\,\rfloor}\,}

\newcommand{\phole}{\square}

\newcommand{\conc}{\;\,}
\newcommand{\emptyseq}{\bullet}

\newcommand{\TSV}{\VV_{\TS}}
\newcommand{\ASV}{\VV_{\AS}}
\newcommand{\VV}{\mathcal{V}}

\allowdisplaybreaks[1]

\newcommand{\st}{simple term}
\newcommand{\sts}{simple terms}
\newcommand{\St}{Simple term}
\newcommand{\Sts}{Simple terms}
\newcommand{\short}{\CalculusShortName}
\newcommand{\ov}[1]{\overline{#1}}

\newif\ifmc
\mcfalse %without color
\newif\ifeg
\title{Hybrid Calculus of Wrapped Compartments\thanks{This research is founded by the BioBITs Project (\emph{Converging
Technologies} 2007, area: Biotechnology-�ICT), Regione Piemonte.}}
\author{Mario Coppo$^1$, Ferruccio Damiani$^1$, Maurizio Drocco$^1$,  Elena Grassi$^{1,2}$, Eva Sciacca$^1$,\\Salvatore Spinella$^1$, Angelo Troina$^1$
\institute{$^1$Dipartimento di Informatica, Universit\`a di Torino}
\institute{$^2$Molecular Biotechnology Center, Dipartimento di Genetica, Biologia e Biochimica, Universit\`a di Torino}
}
\def\titlerunning{Hybrid Calculus of Wrapped Compartments}
\def\authorrunning{M. Coppo, F. Damiani, M. Drocco, E. Grassi, E. Sciacca, S. Spinella and A. Troina}
\begin{document}
\maketitle

\begin{abstract}

The modelling and analysis of biological systems has deep roots in Mathematics, specifically in the field of ordinary differential equations (ODEs). Alternative approaches based on formal calculi, often derived from process algebras or term rewriting systems, provide a quite complementary way to analyze the behaviour of biological systems. These calculi allow to cope in a natural way with notions like compartments and membranes, which are not easy (sometimes impossible) to handle with purely numerical approaches, and are often based on stochastic simulation methods.
Recently, it has also become evident that stochastic effects in regulatory networks play
a crucial role in the analysis of such systems. Actually, in many situations it is necessary to use stochastic models. For example when the system to be described is based on the interaction of few molecules, when we are at the presence of a chemical instability, or when we want to simulate the functioning of a pool of entities whose compartmentalised structure evolves dynamically. In contrast, stable metabolic networks, involving a large number of reagents, for which
the computational cost of a stochastic simulation becomes an insurmountable obstacle, are efficiently modelled with ODEs.
In this paper we define a hybrid simulation method, combining the stochastic approach with ODEs, for systems described in CWC, a calculus on which we can express the compartmentalisation of a biological system whose evolution is defined by a set of rewrite rules.
\end{abstract}

\section{Introduction}

The most common approach of biologists to describe biological systems is based on the use of
deterministic mathematical means like, e.g., ordinary differential equations (ODEs for short). ODEs make it possible to abstractly reason on the behaviour of
biological systems and to perform a quantitative \emph{in silico} investigation. This kind of modelling,
however, becomes more and more difficult, both in the specification phase and in the analysis processes, when
the complexity of the biological systems taken into consideration increases. This has probably been one of the main
motivations for investigating the description of biological systems
by means of formalisms developed in Computer Science for the description of computational entities ~\cite{RS02}.
Different formalisms have either been applied to (or have been inspired
from) biological systems. Automata-based
models~\cite{ABI01,MDNM00} have the advantage of allowing the direct
use of many verification tools such as model checkers. Rewrite
systems~\cite{DL04,P02,BMMT06} usually allow describing
biological systems with a notation that can be easily understood by
biologists. Both automata-like models and rewrite systems
present, in general, problems from the point of view of
compositionality, which allows studying the behaviour of a
system componentwise. Compostionality, instead, is in general ensured by Process calculi, included those commonly used to describe biological systems~\cite{RS02,PRSS01,Car05}. Quantitative simulations of biological models represented with these kind of frameworks (see, e.g.~\cite{PRSS01,DPPQ06,KMT08,BMMTT08,DPR08,preQAPL2010})
are usually developed via a stochastic method derived by Gillespie's algorithm~\cite{G77}.

The ODE description of biological systems determines \emph{continuous}, \emph{deterministic} models in which variables describe the concentrations of the species involved in the system as functions of the time. These models are based on average reaction rates, measured from real experiments which relate to the change of concentrations over time, taking into account the
known properties of the involved chemicals, but possibly abstracting away some unknown mechanisms. Given the reaction equations (together with their rates) and the initial amount for each species, an ODEs model can be constructed by writing a differential equation for each biochemical specie whose concentration changes over time.

In contrast to the deterministic model, \emph{discrete}, \emph{stochastic} simulations involve random variables. Therefore, the behaviour of a reaction is not determined a priori but characterized statistically. Since biological reactions fall in the category of stochastic systems (the very basic steps of every molecular reaction can be described only in terms of its probability of occurrence), stochastic kinetic models are increasingly accepted as the best way to represent and simulate genetic and biochemical networks. Moreover, when the system to be described is based on the interaction of few molecules, or we want to simulate the functioning of a little pool of cells it is necessary to use stochastic models.

The stochastic approach is always valid when the deterministic one is, and it may be valid when the ordinary deterministic is not (i.e. in a nonlinear
system in the neighborhood of a chemical instability).
 Actually, in the last years it has become evident that stochastic effects in regulatory networks play
a crucial role in the analysis of such systems (for example in case of multi-stable systems). In contrast, metabolic
networks involving large numbers of molecules are most often modelled deterministically.
Thus, because of the bimodal nature of biological systems, it may happen that a purely deterministic model does not accurately capture the dynamics of the considered system, and a stochastic description is needed. However,
the computational cost of a discrete simulation often becomes an insurmountable obstacle.
Computationally,  the ODEs method is extremely more efficient. Thus, when the deterministic approach is applicable, it might be profitable to take advantage of
its efficiency, and move to the stochastic approach when it is not. In a hybrid model, some reactions are modelled in a discrete way (i.e. computed, probabilistically, according
to a stochastic method) and others  in a continuous way (i.e. computed, in a deterministic way, by a set of ODEs).

Hybrid models for the simulation of biological systems have been presented in the last few years for purely mathematical models~\cite{SK05,GCPS06,CDR09}. In this paper we adapt the hybrid simulation technique within the programming language approach to describe and analyse the dynamics of biological systems.

In~\cite{preQAPL2010} we proposed the \emph{Calculus of Wrapped Compartments} (\short\ for short), a simplification of the Calculus of Looping Sequences
(CLS for short)~\cite{BMMT06,BMMTT08}. Starting from an alphabet of atomic elements, \short\ terms are defined as multisets of elements and compartments. Elements can be localized by compartmentalisation and the structure of a compartment can be specified by detailing the elements of interest on its
membrane. The evolution of the system is driven by a set of rewrite rules modelling the reactions of interest. We provided \short\ with a stochastic
operational semantics from which a continuous time Markov chain can be build following the standard Gillespie's approach~\cite{G77}.

In this paper we define a hybrid simulation method for systems described in \short ; thus: (1) we are able to simulate systems with compartments, (2) we use the stochastic simulation method when the deterministic one is not valid, and (3) we exploit the efficiency of the deterministic approach whenever it is applicable.

\paragraph{Summary.} Section~\ref{CWC_formalism} introduces the \short\ formalism. Section~\ref{sect:Quant} recalls the stochastic and the deterministic
simulation methods. Section~\ref{sect:Hybrid} introduces the hybrid simulation technique and Section~\ref{sect:tat} applies it to the analysis of the
HIV-1 transactivation mechanism. Finally, in Section~\ref{conc}, we draw our conclusions.

\section{The Calculus of Wrapped Compartments}\label{CWC_formalism}

Like most modelling languages based on term rewriting (notably CLS), a CWC (biological) model consists of a term, representing the system and a set of
rewrite rules which model the transformations determining the system's evolution. The calculus presented here is a slight variant of the one introduced
in~\cite{preQAPL2010}. Namely, compartments are enriched with a nominal type which identifies the set of rewrite rules that can be applied on that
compartment.

\paragraph{\textbf{Terms and Structural Congruence}}%\label{CLS_syntax}

A \emph{term} of the \short\ calculus is intended to represent a biological system. A \emph{term} is a multiset of \emph{\st s}. \Sts, ranged over by $t$,
$u$, $v$, $w$, are built by means of the \emph{compartment} constructor, $(-\into -)^-$, from a set $\AT$ of \emph{atomic elements} (\emph{atoms} for
short), ranged over by $a$, $b$, $c$, $d$, and from a set $\LT$ of \emph{compartments types} (represented as \emph{labels} attached to compartments and
rules), ranged over by $\ell,\ell',\ell_1,\ldots$. The syntax of {\st s} is given at the top of Figure~\ref{fig:CWM-syntax}. We write $\overline{t}$ to
denote a (possibly empty) multiset of \sts\ $t_1\mydots t_n$. Similarly, with $\overline{a}$ we denote a (possibly empty) multiset of atoms. The set of
simple terms will be denoted by $\TT$. The set of terms (multiset of simple terms) and the set of multisets of atoms will be denoted by $\TS$ and $\AS$,
respectively. Note that $\AS \subseteq \TS$.

A term $\overline{t}=t_1\mydots t_n$ should be understood as the multiset containing the simple terms $t_1,\ldots,t_n$. Therefore, we introduce a relation
of structural congruence, following a standard approach in process algebra. The \short\ \emph{structural congruence} is the least equivalence relation on
terms satisfying the rules given  at the bottom of Figure~\ref{fig:CWM-syntax}. From now on we will always consider terms modulo structural
congruence. Then a \st\ is either an atom or a
compartment $(\overline{a}\into \overline{t})^\ell$ consisting of a \emph{wrap} (represented by the multiset of atoms $\overline{a}$), a \emph{content}
(represented by the term $\overline{t}$) and a \emph{type} (represented by the label $\ell$).
%We will identify a \st\ $t$ with the multiset containing only $t$.
%
%The cardinality of a multiset $\overline{t}$ is written $\LengthOf{\overline{t}}$.
We write the empty multiset as $\emptyseq$ and denote the union of
two multisets $\overline{u}$ and $\overline{v}$ as
$\overline{u}\conc\overline{v}$. Let's extend the notion of subset
(denoted as usual as $\subseteq$) between terms interpreted as
multisets.

\begin{figure}%[t]
\hrule $\;$ \\
\textbf{\St s syntax}
\\
 $
 \begin{array}{lcl}
 %\\
  t & \;\qqop{::=}\; & a \!\agr\! (\overline{a}\into\overline{t})^\ell
   \\
  % \\
 \end{array}
 $
\\
%\hrule
$\;$
\\
\textbf{Structural congruence}
\\
$
\begin{array}{l}
%\\
\overline{t} \conc u \conc w \conc \overline{v} \equiv
\overline{t} \conc w \conc u \conc \overline{v} ~~~~~~~~~~~~~~\\
  \mbox{if } \; \ov{a}\equiv\ov{b}~~\text{and}~~\ov{t}\equiv\ov{u} ~~\mbox{ then }~~
    (\ov{a}\into\ov{t})^\ell\equiv(\ov{b}\into\ov{u})^\ell
%~~~~~~~~~~~(\emptyseq\into\emptyseq) \equiv
%\emptyseq~~~~~~~~~~\emptyseq\, u \equiv u \,\emptyseq \equiv u
%
%\\
%
%\mbox{if }
%\quad \overline{a} \equiv \overline{b}  \mbox{ and } \overline{t} \equiv \overline{u}
%\quad \mbox{ then } \quad
%\overline{v} \conc (\overline{a}\into\overline{t}) \overline{w}
%\equiv
%\overline{v} \conc (\overline{b}\into\overline{u}) \conc \overline{w}
\\
%\\
\end{array}
$ %\hrule
 \caption{ \CalculusShortName\ term syntax and structural congruence rules}
\label{fig:CWM-syntax}
\end{figure}

\begin{figure*}%[t]
\begin{center}
\begin{minipage}{0.98\textwidth}
\begin{center}
\includegraphics[height=30mm]{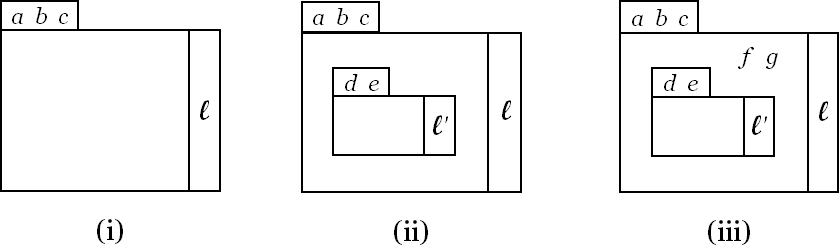}
\end{center}
\vspace{-0.3cm} \caption{\textbf{(i)} represents $(a \conc b \conc c \into \emptyseq)^\ell$; \textbf{(ii)} represents $(a \conc b \conc c  \into (d \conc e \into
\emptyseq)^{\ell'})^\ell$; \textbf{(iii)} represents $(a \conc b \conc c \into (d \conc e \into \emptyseq)^{\ell'} \conc f \conc g)^\ell$} \label{fig:example CWM}
\end{minipage}
\end{center}
\end{figure*}

An example of term is $\ov{t} = a \conc b \conc (c \conc d \into e \conc f)^\ell$ representing a multiset consisting of two atoms $a$ and $b$ (for instance two
molecules) and an $\ell$-type compartment $(c \conc d \into e \conc f)^\ell$ which, in turn, consists of a wrap (a membrane) with two atoms $c$ and $d$
(for instance, two proteins) on its surface, and containing the atoms $e$ (for instance, a molecule) and $f$ (for instance a DNA strand). See
Figure~\ref{fig:example CWM} for some graphical representations.

%%%%%FIGURA

\paragraph{\textbf{Rewrite Rules, Variables, Open Terms and Patterns}}
A rewrite rule is defined as a pair of terms (possibly containing variables), which represent the patterns defining the system transformations, together
with a label $\ell$ representing the compartment type on which the rule can be applied. Compartments are identified by the notion of (labelled)
reduction context introduced below.  A rule is applicable in a compartment if its content matches the left-hand side of the rule via a proper instantiation of its variables (note this instantiation is in general not unique). A system transformation is obtained by replacing the reduced subterm by the corresponding instance of the right-hand side of the rule.

%%%%%%%%%%%%%%%%%%%%%%%%%%%%%%%%%

In order to formally define the rewriting semantics, we introduce
the notion of open term (a term containing variables) and pattern
(an open term that may be used as left part of a rewrite rule). In
order to respect the syntax of terms, we distinguish between
``wrap variables'' which may occur only in compartment wraps (and
can be replaced only by multisets of atoms) and ``term variables''
which may only occur in compartment contents or at top level (and
can be replaced by arbitrary terms).
Therefore, we assume a set of \emph{term variables}, $\TSV$, ranged over by $X,Y,Z$, and a set of \emph{wrap variables}, $\ASV$, ranged over by $x,y,z$.
These two sets are disjoint. We denote by $\VV$ the set of all variables $\TSV \cup \ASV$, and with $\rho$ any variable in $\VV$.

\begin{itemize}
\item \emph{Open terms} are terms
    which may contain occurrences of wrap variables in compartment
    wraps and term variables in compartment contents or at top level.
    They can be seen as multisets of \emph{simple open terms}.
    More formally, open terms, ranged over by $\RightPat$ and simple open terms, ranged over by $o$, are defined in the following way:
    $$
\begin{array}{lcl}
\RightPat   & \; \qqop{::=} \; & \overline{o} \\
o           & \; \qqop{::=} \; & a \agr X \agr   (\overline{a}\conc\overline{x}  \into \overline{o} )^\ell
\end{array}
$$
We denote with $\RightPP$ the set of open terms. An open term is \emph{linear} if each variable occurs at most
    once.
%\item An open term is a multiset of open \sts.
 \item \emph{Patterns}, ranged over by $\LeftPat$, and \emph{simple patterns}, ranged over by $\Pat$, are the linear open terms defined in the following way:
% \small
$$
\begin{array}{lcl}
  \LeftPat & \;\qqop{::=}\; & \overline{\Pat} \\
  \Pat & \;\qqop{::=}\; & t \agr (\overline{a} \conc x \into \overline{\Pat}\conc X)^\ell
\end{array}
$$
%\normalsize where $t$ is an element of $\TT$, $\overline{a}$ is an
%element of $\AS$, $x$ is a variable in $\ASV$, $\overline{\Pat}$
%is a possibly empty multiset of simple patterns and $X$ is a
%variable in $\TSV$.
We denote with $\LeftPP$ the set of patterns.
 \end{itemize}

An \emph{instantiation} is a partial function $\sigma : \VV
\rightarrow \TS$. An instantiation must preserve the type of
variables, thus for $X \in \TSV$ and $x \in \ASV$ we have
$\sigma(X) \in \TS$ and $\sigma(x) \in \AS$, respectively.  Given
$\RightPat \in \OT$, with $\RightPat \sigma$ we denote the term
obtained by replacing each occurrence of each variable $\rho \in
\VV$ appearing in $\RightPat$ with the corresponding term
$\sigma(\rho)$.

Let $\Sigma$ denote the set of all the possible instantiations and
$\VarOf(\RightPat)$ denote the set of variables appearing in
$\RightPat\in \OT$.

A \emph{rewrite rule} is a triple $(\ell,P,\RightPat)$, also denoted by $\ell:\LeftPat \! \srewrites{} \! \RightPat$, where $\LeftPat \in \LeftPP$ and
$\RightPat \in \OT$ are such that $\VarOf(\RightPat) \subseteq \VarOf(\LeftPat)$. The label $\ell$ denotes the type of the compartments where the rule can
be applied. A rewrite rule $\ell:\LeftPat \! \srewrites{} \! \RightPat$ then states that a subterm $P \sigma$, obtained by instantiating variables in
$\LeftPat$ by some instantiation function $\sigma$, can be transformed into the subterm $\RightPat\sigma$ within any compartment of type $\ell$. We use
the special label $\TOP\in\LT$ to denote the type of the top level of a term.

\paragraph{\textbf{Contexts}}
The definition of reduction for CWC systems is completed by
resorting to the notion of reduction context. To this aim, the syntax of terms is enriched with a new element $\phole$
representing a hole. \emph{Reduction context} (ranged over by $C$)
are defined by:
$$
 C  \;\qqop{::=}\; \phole \agr C \conc \overline{t} \agr (\overline{a} \into C)^\ell
$$
where $\overline{a} \in \AS$, $\overline{t}\in \TS$ and $\ell \in \LT$. We denote with $\CC$ the infinite set of contexts.

By definition, every context contains a single hole $\phole$. Let us assume $C,C'\in \CC$. With $C[\overline{t}]$ we denote the term obtained by replacing
$\phole$ with $\overline{t}$ in $C$; with $C[C']$ we denote context composition, whose result is the context obtained by replacing $\phole$ with $C'$ in
$C$. For example, given $C=(a \conc b \into \phole)^\ell\conc  i $, $C'= (c\conc d \into \phole)^{\ell'}\conc  g \conc h$ and $\overline{t}= e \conc f$,
we get $C[C'[\overline{t}]]= (a \conc b \into (c \conc d \into e \conc f)^{\ell'} \conc g \conc h )^\ell\conc  i $.

%Note that context
%holes take the place either of the whole term or of the whole
%content of a compartment. This allows to make context unambiguous
%in the following sense:
%\begin{proposition}[Uniqueness]\label{unicity}
%For any term $\ov{t}$ if the term $\ov{u}$  occurs in $\ov{t}$
%within a compartment content or at top level, then there are,
%modulo $\equiv$, a unique context $C$ and a unique term $\ov{t'}$
%such that $\ov{t} = C[\ov{t'}]$ and $\ov{u}\subseteq \ov{t'}$.
%\end{proposition}
%\\

In order to apply a rule within a compartment of the correct type we define a function that, given a context, returns the label of the innermost
compartment containing the hole. If the hole appears at top level, the distinguished label $\TOP$ is returned. The function $\LAB$ is defined as follows:
$$
\LAB(C) = \left\{
\begin{array}{lcl}
\TOP & \quad & \textrm{ if } C = \phole \conc \overline{t} \\
\ell    & \quad & \textrm{ if } C =C'[(\overline{a} \into \phole \conc \overline{t})^\ell ]
\end{array} \right.
$$

\paragraph{\textbf{Qualitative Reduction Semantics}}
A \emph{\short\ system} over a set $\AT$ of atoms and a set $\LT$ of labels is represented by a set $\QQ_{\AT,\LT}$ ($\QQ$ for short when $\AT$ and $\LT$
are understood) of rewrite rules over $\AT$ and $\LT$.

 The \emph{qualitative reduction semantics} of a \short\ system $\QQ$ is the least transition relation satisfying the following
rule:
$$
\frac{ \ell:P \srewrites{}  \RightPat \in \QQ \quad\sigma \in \Sigma
           \quad C\in \CC \quad \LAB(C)=\ell}
    { C[P\sigma] \ltrans{}  C[O\sigma]}
$$

\paragraph{\textbf{Modelling Guidelines}}%\label{sec:modelguide}
In this section we  give some explanations and general hints about how \short\ could be used to represent the behaviour of various biological systems.
Here, entities are
represented by terms of the rewrite system, and events by
rewrite rules.

First of all, we should select the biomolecular entities of
interest. Since we want to describe cells, we consider molecular
populations and membranes. Molecular populations are groups of
molecules that are in the same compartment of the cells and inside
them. As we have said before, molecules can be of many types: we
classify them as proteins, chemical moieties and other molecules.
%One could describe these molecules at different levels: for example
%proteins can be seen as amino acids sequences, domains sequences or
%as a single entity. In \CalculusShortName{ }the simplest choice
%is to use a single symbol to represent a protein.

Membranes are considered as elementary objects: we do not describe
them at the level of the phospholipids they are made of. The only
interesting properties of a membrane are that it may have a
content (hence, create a compartment) and that in its phospholipid
bilayer various proteins are embedded, which act for example as
transporters and receptors. Since membranes are represented as
multisets of the embedded structures, we are modeling a fluid
mosaic in which the membranes become similar to a two-dimensional
liquid where molecules can diffuse more or less
freely~\cite{SN72}.

Compartment labels are useful to identify the kind of a compartment. For example, we may use compartment labels to denote a nucleus within a cell, the different organelles, etc..

Table~\ref{tab:guidelines-events} lists the guidelines (taken from~\cite{preQAPL2010}) for the abstraction into CWC rules of some basic biomolecular events, some of which will be used in our applications.\footnote{Compartment labels are omitted for simplicity, just notice that the rules shown in the table can be specified to
apply only within a given type of compartment.} Entities are associated with CWC terms: elementary objects (genes, domains, etc...) are modelled as atoms,
molecular populations as CWC terms, and membranes as atom multisets. Biomolecular events are associated with CWC rewrite rules.

%A selection of common biomolecular event sismulation .
The simplest
kind of event is the change of state of an elementary object.
Then, there are interactions between molecules: in particular
complexation, decomplexation and catalysis. Interactions could
take place between simple molecules, depicted as single symbols,
or between membranes and molecules: for example a molecule may
cross or join a membrane. Finally, there are also interactions
between membranes: in this case there may be many kinds of
interactions (fusion, vesicle dynamics, etc\ldots).

\begin{table}%[t]
\begin{center}
\begin{footnotesize}
\begin{tabular}{|l|l|}
\hline
{\bf Biomolecular Event} & {\bf \short\ Rewrite Rules} \\
\hline \hline State change &
    $a \srewrites{} b$ \\
\hline Complexation &
    $a \conc b \srewrites{} c$ \\
\hline Decomplexation &
    $c \srewrites{} a \conc b$ \\
\hline State change on membrane &
    $ (a \conc x \into X) \srewrites{} (b \conc x \into X) $ \\
\hline Complexation &
    $ a \conc (b \conc x \into X) \srewrites{}  (c \conc x \into X) $ \\
on membrane
    &  $ (b \conc x \into a \conc X) \srewrites{}  (c \conc x \into X) $ \\
\hline Decomplexation
    &  $ (c \conc x \into X) \srewrites{}  a \conc (b \conc x \into X) $ \\
on membrane
    &  $ (c \conc x \into X) \srewrites{}  (b \conc x \into a \conc X) $ \\
\hline Membrane crossing &
    $ a \conc (x \into X) \srewrites{}  (x \into a \conc X) $\\
    & $ (x \into a \conc X) \srewrites{}  a \conc (x \into X) $\\
\hline
Catalyzed&  $ a \conc (b \conc x \into X) \srewrites{}  (b \conc x \into a \conc X) $\\
membrane crossing & $ (b \conc x \into a \conc X) \srewrites{}  a \conc (b \conc x \into X) $\\
\hline Membrane joining &
    $ a \conc (x \into  X) \srewrites{}  (a \conc x \into X) $\\
    & $ (x \into a \conc X) \srewrites{} (a \conc x \into X)$\\
\hline Catalyzed &
    $ a \conc (b\conc x \into X) \srewrites{} (a \conc b \conc x \into X) $ \\
membrane joining &
    $ (b\conc x \into a \conc X) \srewrites{} (a \conc b \conc x \into X) $ \\
    & $ (x \into a \conc b \conc X) \srewrites{}(a \conc x \into b \conc X) $\\
\hline
\end{tabular}
\end{footnotesize}
\end{center}
\caption{Guidelines for modelling biomolecular events in \short}\label{tab:guidelines-events}
\end{table}

\section{Quantitative Simulation Models for \short}\label{sect:Quant}

A quantitative operational semantics for \short\ can be defined by associating to the rewriting rules of \short\ the kinetic constant $k$ of the modeled
chemical reaction. A \emph{quantitative rewrite rule} is then a quadruple $(\ell,P,\RightPat, k)$, denoted with $\ell:\LeftPat \! \srewrites{k} \!
\RightPat$, where $\ell,P$ and $\RightPat$ are in Section~\ref{CWC_formalism}, and $k \in \bbbr^{\geq 0}$.

In this section we introduce two (standard) quantitative simulation methods for CWC based respectively on Gillespie's stochastic simulation algorithm
\cite{G77} and on the (deterministic) solution of ordinary differential equations. These two approaches, that will be presented separately in this section,
will be integrated in the next section defining the hybrid semantics of CWC.

A prototype implementation of the hybrid CWC calculus (which encompasses both the pure stochastic and the deterministic versions of the calculus) is
available~\cite{HCWC_SIM}.

\begin{remark}
Notice that the stochastic, deterministic and hybrid approaches introduced in this section simulate well-stirred system of molecules, confined to a constant volume and in thermal equilibrium at some constant temperature. In these conditions we can describe the system's state by specifying only the molecular populations, ignoring the positions and velocities of the individual molecules.
Different approaches such as Molecular Dynamics, Partial Differential Equations or Lattice-based methods are required in case of molecular crowding, anisotropy of the medium or canalization.
\end{remark}

\paragraph{\textbf{Running Example}}
In order to illustrate the quantitative semantics of \short\ we consider, as a running example, a toy case study based on a compartmentalized variant of a system studied in ecology to
describe a generalized competitive Lotka-Volterra dynamics~\cite{SM76}. The variant consists in an schema of ecoregions bounded by geographical frontiers rendered by compartments. The system has $3$ competitive species in $2$ environmental compartments. The
supposed population dynamics depend on the following parameters:
\begin{itemize}
  \item  \emph{the interaction matrix} $\mu_{i,j} \geq 0$,
  where the element $\mu_{i,j}$ represents the relative strength that species $i$ has on the population of species $j$, i.e. the competition between species in the same environment due to incompatibility;
  \item \emph{the carrying capacity} $K_i \geq 0$ is the population size of the species $i$ that the environment can sustain indefinitely assuming no interaction between the species;
  \item \emph{the migration rate} $d_i$ associated to each species which migrates between the compartments.
\end{itemize}

\begin{table}%[t]
\begin{center}
\begin{tabular}{c|c|c|c}
$\mu$ & \textbf{A} & \textbf{B} & \textbf{C} \\
\hline
\textbf{A} & 0.15 & 0.2 & 0.15 \\
\hline
\textbf{B} & 0.2 & 0.15 & 0.2 \\
\hline
\textbf{C} & 0.15 & 0.2 & 0.15 %\\
\end{tabular}
\end{center}
\caption{Interaction Matrix}
\label{inter_matrix}
\end{table}

In our tests, we set the interaction matrix in accordance to Table~\ref{inter_matrix}. The carrying capacities are $K_i = 100$ and the migration rates
between the compartments are: $d_A = 0.01$, $d_B = 0.01$, $d_C = -0.01$. The migration rates are positive for species $A$ and $B$ moving from the
compartment to the outside, and negative for species $C$ which moves in the reverse path.
This system has a wide set of possible behaviour, particularly the compartmentalization can be interpreted as the case of an ecological frontier like a river or a
mountain which partially separates the different populations.

\begin{figure}%[t]
\hrule
$
\begin{array}{lc}
(N_1) &  \TOP : (x \into A \conc X)^{\textit{IN}} \srewrites{d_A} (x \into X)^{\textit{IN}} \conc A \\% \label{Aexport}\\
(N_2) &   \TOP : (x \into B \conc X)^{\textit{IN}} \srewrites{d_B} (x \into X)^{\textit{IN}} \conc B \\% \label{Bexport}\\
(N_3) &   \TOP : (x \into X)^{\textit{IN}} \conc C \srewrites{- d_C} (x \into C \conc X)^{\textit{IN}}\\% \label{Cimp}\\
(B_1) &   \TOP, \textit{IN} : A \srewrites{1}  A \conc A \\ % \label{Aprod}\\
(B_2) &  \TOP, \textit{IN} : B \srewrites{1}  B \conc B \\ % \label{Bprod}\\
(B_3) & \TOP, \textit{IN} : C \srewrites{1}  C \conc C \\ % \label{Cprod}\\
(B_4) & \TOP, \textit{IN} : A \conc A \srewrites{k_{A,A}}  \emptyseq \\ % \label{Adegr}\\
(B_5) & \TOP, \textit{IN} : B \conc B \srewrites{k_{B,B}}  \emptyseq \\ % \label{Bdegr}\\
(B_6) & \TOP, \textit{IN} : C \conc C \srewrites{k_{C,C}}  \emptyseq \\ % \label{Cdegr}\\
(B_7) & \TOP, \textit{IN} : A \conc B \srewrites{k_{A,B}}  \emptyseq \\ % \label{ABdegr}\\
(B_8) &\TOP, \textit{IN} : A \conc C \srewrites{k_{A,C}}   \emptyseq \\ % \label{ACdegr}\\
(B_9) & \TOP, \textit{IN} : B \conc C \srewrites{k_{B,C}}  \emptyseq % \label{Cdegr}
\end{array}
$
 \caption{ \CalculusShortName\ rules for the test case}
\label{fig:TOY-rules}
\end{figure}

The set of \short\ rules adopted in our toy case study is given in Figure~\ref{fig:TOY-rules}, where the rates of competition of the species $i$ against the species $j$ were calculated as
$k_{i,j}=\frac{\mu_{i,j}}{K_i}$. The two environmental compartments are represented by the implicit top level compartment (of type $\TOP$) and by an
explicit compartment of type \textit{IN}. Rules $(N_1 - N_3)$ model the migration of the three species between the two compartments. The
other rules model the competition between species (rules $B_4 - B_9$) and their reproduction capacity (rules $B_1 - B_3$). Note that rule $(B_1)$, is
 indeed a compact representation for the rules:
 $$
\begin{scriptsize}
\begin{array}{c@{\qquad\qquad\qquad}c}
(B'_1) \;\;\;   \TOP : A \srewrites{1}  A \conc A
  &
(B''_1) \;\;\;   \textit{IN} : A \srewrites{1}  A \conc A
\end{array}
\end{scriptsize}
$$
Similarly for the rules $(B_2),\ldots,(B_9)$.
The simulations will be performed for $35$ time units, with the starting term:\footnote{The notation $n \times a$, where $n$ is a natural number and $a$
is an atom, denotes the multiset containing $n$ occurrences of $a$.}
$$
2 \times C \conc (\emptyseq \into 2 \times A\conc 2 \times B)^{\textit{IN}}.
$$

\subsection{Stochastic Evolution}\label{SECT:STO_SEM}

A stochastic simulation model for biological systems can be defined by incorporating a collision-based stochastic framework along the line of the one
presented by Gillespie in \cite{G77}, which is, \emph{de facto}, the standard way to model quantitative aspects of biological systems. The idea of
Gillespie's algorithm is that a rate constant is associated with each considered chemical reaction. Such a constant is obtained by multiplying the kinetic
constant of the reaction by the number of possible combinations of reactants that may occur in the system. The resulting rate is then used as the
parameter of an exponential distribution modelling the time spent between two occurrences of the considered chemical reaction. Following the law of mass
action, it is necessary to count the number of reactants that are present in a system in order to compute the exact rate of a reaction. The same approach
has been applied, for instance, to define the quantitative semantics of the stochastic $\pi$-calculus \cite{P95,PRSS01}.

The use of exponential distributions to represent the (stochastic)
time spent between two occurrences of chemical reactions allows
describing the system as a Continuous Time Markov Chain (CTMC), and
consequently allows verifying properties of the described system
analytically and by means of stochastic model checkers.

%\subsubsection{The Stochastic Reduction Semantics for \short }\label{SECT:STOCH_SEM}

The number of reactants in a reaction represented by a rewrite rule is evaluated considering the number of distinct occurrences, in the same context, of
subterms to which the rule can be applied producing the same term. For instance in evaluating the application rate of the stochastic rewrite rule $R=
\TOP: a \conc b \srewrites{k} c$ to the term $\ov{t}=a\conc a \conc b \conc b$ we must consider the number of the possible combinations of reactants of
the form $a\conc b$ in $\ov{t}$. Since each occurrence of $a$ can react with each occurrence of $b$, this number is 4. So the application rate of $R$ is
$k\cdot 4$.

The evaluation of the application rate of a reduction rule containing variables is more complicate since there can be many different ways in which
variables can be instantiated to match the subterm to be reduced, and this must be considered to correctly evaluate the application rate. Given two terms
$\ov{t},\ov{u}$ and a reduction rule $R$ we can compute the number of possible applications of the rule $R$ to the term $\ov{t}$ in a context $C[\;]$ of type $\ell$, resulting in the term
$C[\ov{u}]$. We denote this number by $\OO(R, C[\ov{t}],C[\ov{u}])$, where the function $\OO$ is analogous to the one defined for SCLS in~\cite{BMMTT08}.
The \emph{stochastic reduction semantics} of a \short\ system $\QQ$ is the least labelled transition relation satisfying the following rule:
$$
\frac{ R=\ell:P \srewrites{k}  O \in \QQ \quad \sigma \in \Sigma
        \quad C\in \CC \quad \LAB(C)=\ell }
    { C[P\sigma] \ltrans{ k \cdot \OO(R, C[P\sigma],C[O\sigma])}  C[O\sigma]}
$$
Reductions determined by a rule $R$  are labelled with their rates. The rate of a reduction is obtained as the product of the rewrite rate
constant and the number of occurrences of the rule within the starting term (thus counting the exact number of reactants to which the rule can be applied
and which produce the same result). The rate associated with each transition in the stochastic reduction semantics is the parameter of an exponential
distribution that characterizes the stochastic behaviour of the activity corresponding to the applied rewrite rule. The stochastic semantics is
essentially a \emph{Continuous Time Markov Chain} (CTMC).
Given a term $\ov{t}$, a global time $\delta$ and all the reductions $e_1,\ldots,e_M$ that can be applied to $\ov{t}$, with rates $r_1,\ldots,r_M$ such that $r=\sum_{i=1}^M
r_i$, the standard simulation
procedure that corresponds to Gillespie's simulation algorithm~\cite{G77} consists of the following two steps:

\begin{enumerate}
\item The time $\delta+\tau$ at which the next stochastic reduction will occur is randomly
chosen with $\tau$ exponentially distributed with parameter $r$;
\item The reduction $e_i$ that will occur at time $\delta+\tau$ is randomly
chosen with probability $\frac{r_i}{r}$.
\end{enumerate}

\paragraph{\textbf{Running Example: Stochastic Simulations}}

\begin{figure*}[t]
\centering
%\subfigure[First experiment]{
  \includegraphics[width=.45\textwidth]{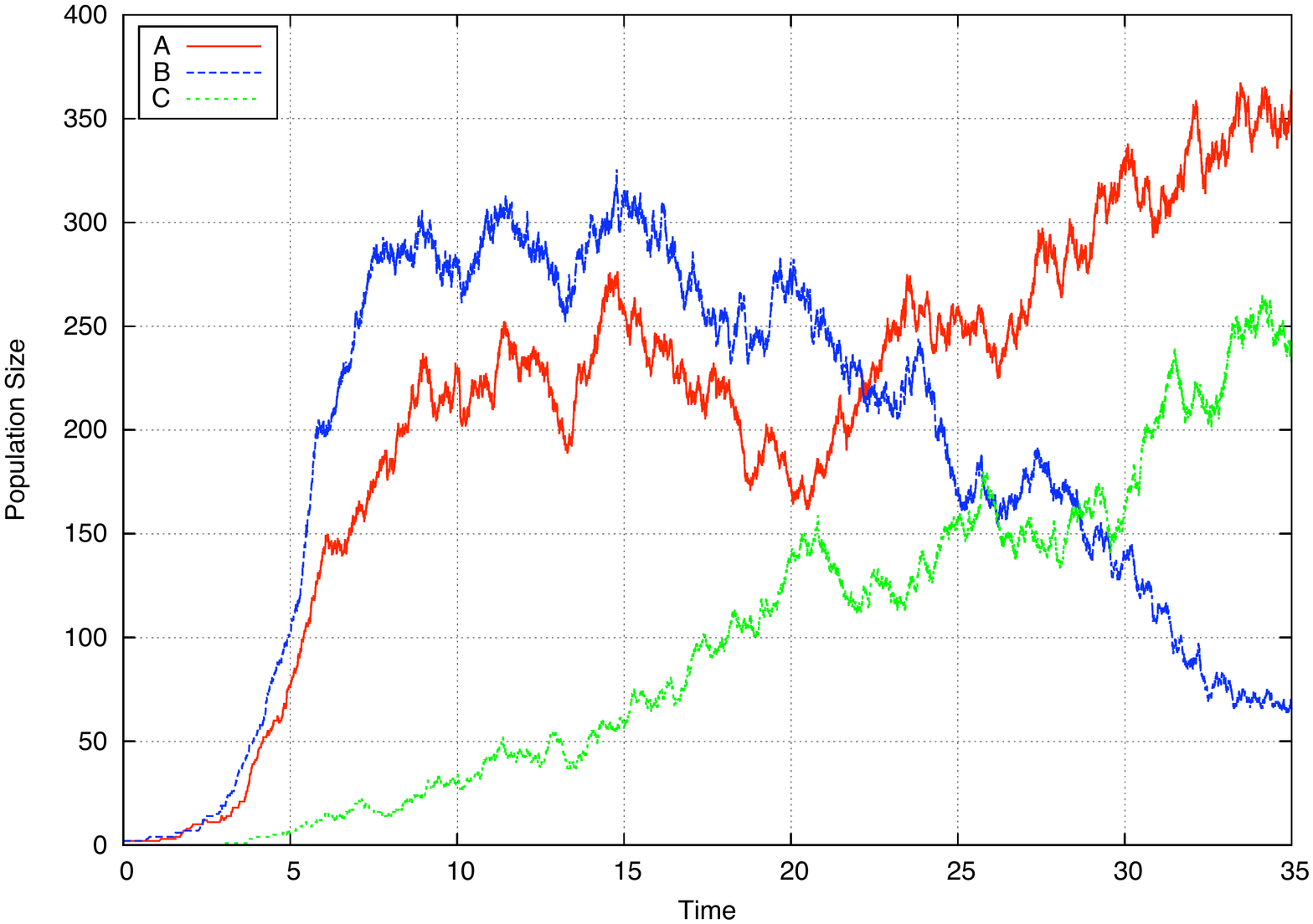}
 % \label{stochINsim}}
%\centering
%\subfigure[Inside the compartment IN]{
 \includegraphics[width=.45\textwidth]{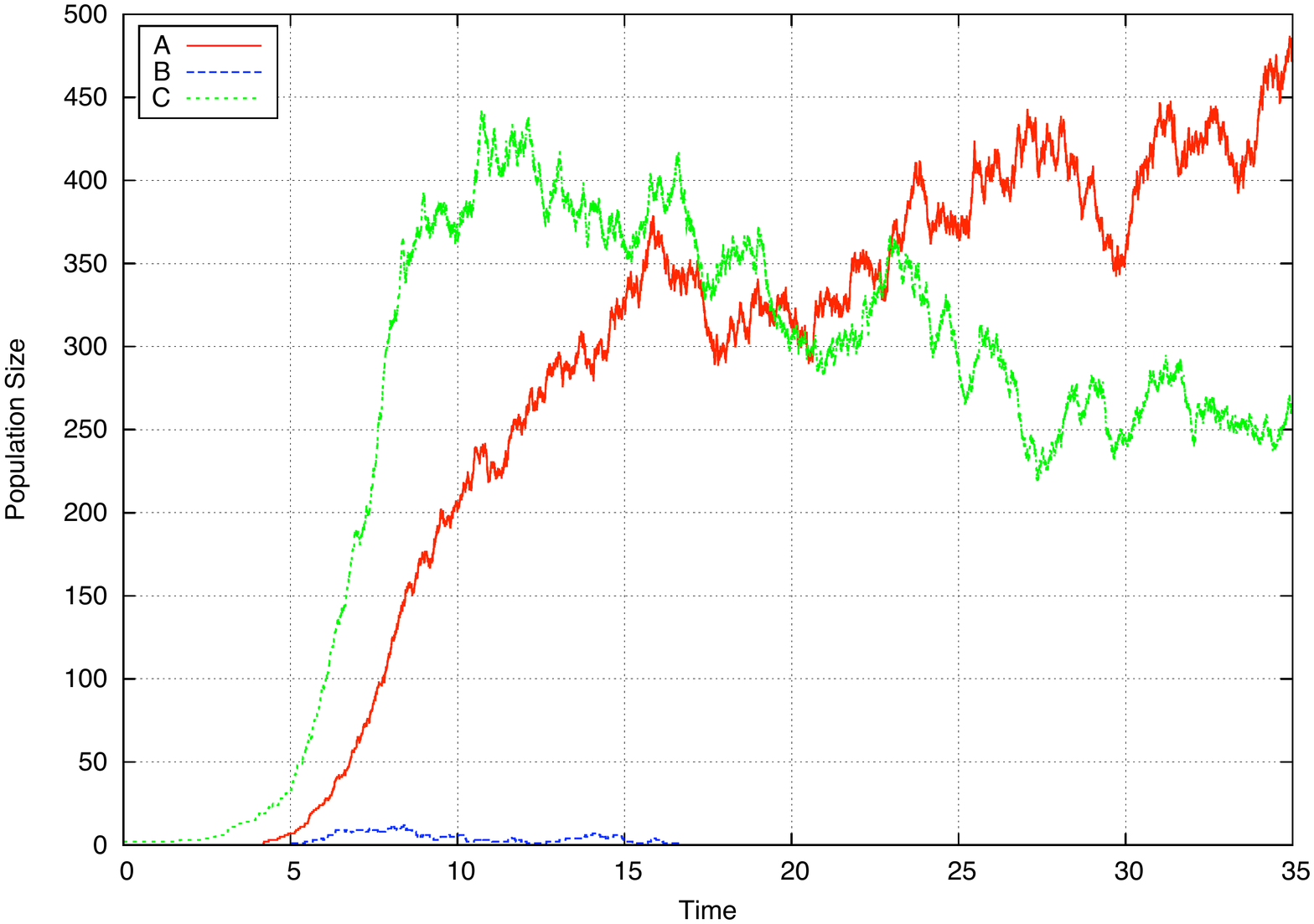}
%  \label{stochINdissimilar1}
%}
\hrule
%\centering
%\subfigure[Second experiment]{
  \includegraphics[width=.45\textwidth]{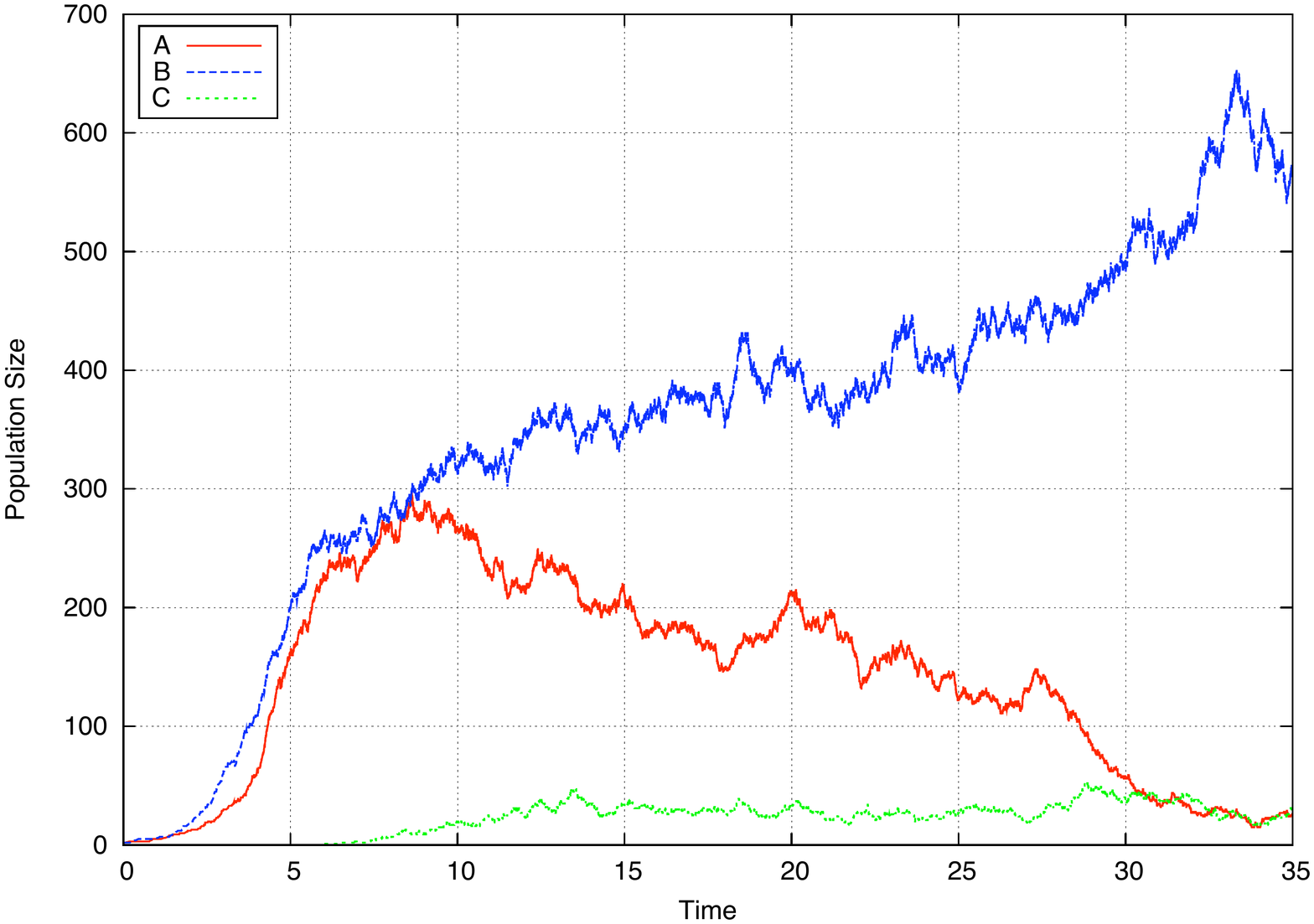}
%  \label{stochTopLevelsimilar}}
%\centering
%\subfigure[In $\intercal$]{
  \includegraphics[width=.45\textwidth]{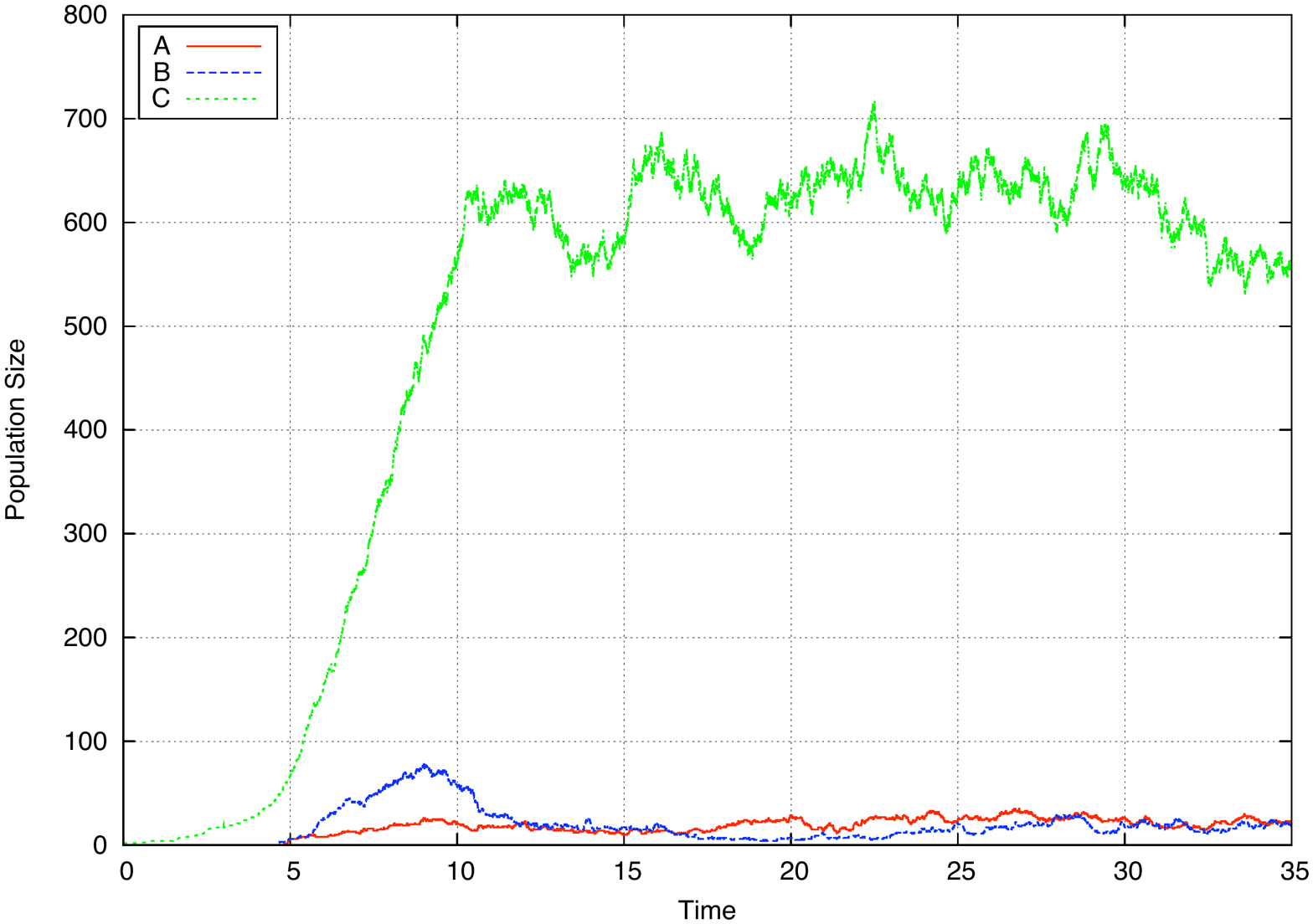}
 % \label{stochTopLeveldissimilar}
 %}
  \caption{Two different runs of the stochastic simulations showing the different behaviour of the dynamics of the competitive species
  inside the compartment \textit{IN} (on the left side) and outside the compartment (on the right side)}
\label{fig:stoch_toy}
\end{figure*}

We performed several stochastic simulations of the toy case study, showing many, different, possible evolutions. Two of these runs are shown in Figure \ref{fig:stoch_toy}.

A characteristic of this example is that in the initial phase, mainly due to the small number of individuals involved, the evolution of the system is
strongly determined by random events that can change dramatically the destiny of the species. In the first experiment, on the top of Figure \ref{fig:stoch_toy},
%the behaviour pattern is similar to the deterministic one which is shown in Figure \ref{fig:ode_toy}. In fact, in this case,
the populations $A$ and $C$ overtake population $B$ both inside and outside compartment \textit{IN}. The second experiment, on the bottom of Figure \ref{fig:stoch_toy}, shows a completely different fate for the populations: namely, population $B$ overtakes populations $A$ and $C$ inside the compartment \textit{IN} while population $C$ overtakes populations $A$ and $B$ outside the compartment.
Note
that %the probabilistic fluctuations underneath the variability of our stochastic simulations are driven by a chain of random events:
the cases shown here are just two possible examples of the many different destinies for the three populations. The second one, in particular, differs sensibly from the average behaviour appearing in the deterministic simulation which is shown in Figure \ref{fig:ode_toy}.

%A characteristic of this example is that in the first phase, owing to the small number of individuals involved, the evolution of the system is strongly determined by casual events, that can change dramatically the destiny of the species. After a while, when the number of individual increases, the statistical fluctuations have an increasingly lower influence on the overall behaviour of the system.

\subsection{Deterministic Evolution}\label{SECT:ODE_SEM}

The standard way to express the evolution of a biochemical system is via ODEs.
%This approach is standard and computationally simple and fast, but has some well known limitations as remarked below.
%
We define the deterministic reduction semantics for a subset of \short\ quantitative rewrite rules, that we call biochemical rewrite rules,
expressing biochemical reactions.

%\subsubsection{Biochemical Patterns  and \short}

%Coupled biochemical reactions are expressed in \short\ with a simplified form of pattern.

%\begin{definition}
%\emph{Biochemical patterns}, ranged over by $\BPat$, are the patterns (elements of $\LeftPat$) defined in the following way:
%\begin{align*}
% \BPat  & \; \qqop{::=} \; \overline{a}
%\end{align*}
%where $\overline{a}$ is a multiset of atomic elements. We denote with $\BPP$ the set of biochemical patterns.
%\end{definition}
%
\emph{Biochemical rewrite rules} are the quantitative rewrite rules of the form $\ell:\overline{a} \srewrites{k} \overline{b}$, where $\overline{a}$ and
$\overline{b}$ are multisets of atomic elements.
%We denote with $\BB$ the set of biochemical reactions.

%\begin{definition} A \emph{coupled biochemical reaction}, occurring in a compartment of type $\ell$, can be expressed in \short\ with a
%rewrite rule $\ell:\overline{a} \srewrites{k} \overline{b}$ between biochemical patterns.
%\end{definition}

All the reactants of a biochemical reaction are completely specified, since both sides of the rules do not contain variables. Moreover,  biochemical
reactions are local to a single compartment. Reactions that invoke and/or change the structure of compartments cannot be expressed with biochemical
rewrite rules. Actually, referring to Table~\ref{tab:guidelines-events}, we notice that biochemical rewrite rules can be used to model state change,
complexation and decomplexation: these are exactly the kinds of reactions naturally eligible to be simulated with ODEs.

A \short\ system $\QQ$ consisting of $r$ biochemical rewrite rules represents
 a system of $r$ biochemical reactions. Its deterministic semantics is defined by
extracting from $\QQ$ a system of ODEs  to be used for simulating the evolution of the involved multisets of atoms.
%
%\subsubsection{ODEs for Biochemical Patterns in \short}
%
For every label $\ell$, let
\begin{itemize}
\item
$a_1,\ldots,a_{n_\ell}$ ($n_\ell\ge 1$) denote the $n_\ell$ species of atoms that may occur at top level within a compartment of type $\ell$, and
\item
$\QQ_\ell$ denote the set of rules with label $\ell$.
\end{itemize}
The $i$-th rule in the set $\QQ_\ell$ is denoted by
$$
  \ell :  \bar{a}_i \srewrites{k_i}  \bar{b}_i \ \ \  i= 1, 2, \ldots, \LengthOf{\QQ_\ell}
$$
 For all species $a_j$ ($j=1, 2, \ldots, n_\ell$) let $\alpha^-_{i,j}$ be the number of atoms of species $a_j$ consumed by the $i$-th rule
 and $\alpha^+_{i,j}$ the number of atoms of species $a_j$ produced by the $i$-th rule.
 The $n_\ell \times \LengthOf{\QQ_\ell}$ stoichiometric matrix $\Lambda_\ell$ is defined
 by $\nu_{i,j}=\alpha^+_{i,j}-\alpha^-_{i,j}$.\footnote{Many of the $\alpha^-_{i,j},\;\alpha^+_{i,j}$ are usually $0$.}

Let $[a]$ denote the concentrations of the atoms of specie $a$ occurring at top level in a given compartment of type $\ell$. If $\bar{a}_i =
a_{i_1}\;\ldots\;a_{i_{r_i}}$ ($r_i\ge 1$), let $[\bar{a}_i]$
 denote the product $[a_{i_1}]\;\ldots\;[a_{i_{r_i}}]$ of the concentrations of the species occurring in $\bar{a}_i$ in the considered
compartment.\footnote{Being $\bar{a}_i$ a multiset, if an element $a_j$ occurs h times in $\bar{a}_i$ (for instance, for $h=2, i_p = i_q$ for some $(1
\leq p,q \leq i_{r_i},\;p\neq q)$ its concentration is considered $h$ times in $[\bar{a}_i]$, i.e. we take the h-th power of $[a_j]$.}
%while $\bar{a}^i \subseteq S$ and $\bar{b}^i \subseteq S$ represents the species involved
%in the $i$-th reaction as reactants and as products.
%
The evolution of the given compartment of type $\ell$ is modelled by the following system of ODEs:
$$
  \ell : \frac{d[s_j]}{dt} = \sum_{i=1}^{\LengthOf{\QQ_\ell}} \nu_{i,j} \cdot k_i \cdot [\bar{a}_i]
$$

%\begin{equation}\label{reaction01}
%  \ell :  \bar{a}^i \rightarrow^{k_i}  \bar{b}^i \ \ \  i= 1, 2, \ldots, M_\ell
%\end{equation}
%the $i$-th reaction in the set $M_\ell$, where $k_i$ is the kinetic rate of the $i$-th reaction while $\bar{a}^i \subseteq S$ and $\bar{b}^i \subseteq S$ represents the species involved in the $i$-th reaction as reactants and as products.
%
%We denote the $j$-th component of $\bar{a}^i$ as $a_j^i$. For all species $j=1, 2, \ldots, N_\ell$ let $\alpha_{i,j}$ be the number of molecules of specie $a_j^i$ consumed by the $i$-th reaction and $\beta_{i,j}$ the number of molecules of species $b_j^i$ produced by the $i$-th reaction. The $N_\ell \times M_\ell$ stochiometric matrix $\Lambda_\ell$ is defined by $\nu_{i,j}=\beta_{i,j}-\alpha_{i,j}$.
%
%To model the evolution of the compartment $\ell$ we will use the following system of Ordinary Differential Equations (ODEs) :
%
%\begin{equation}\label{ode01}
%  \ell : \frac{d[s_j]}{dt} = \sum_{i=1}^{M_\ell} \nu_{i,j} \cdot k_i \cdot \prod_{s_h^i \in \bar{a}^i} [s_h^i] \ \ \ \ j=1,2,\ldots,N_\ell
%\end{equation}
    %
%where $[s_j]$ represents the concentration of the species $s_j \in S$ for $j=1,2,\ldots,N_\ell$.

Computationally, ODEs are well studied and understood. They can be solved
using a variety of numerical methods, from the Euler method to
higher-order Runge-Kutta methods or stiff methods, many of which are readily
available in software packages that can be easily incorporated
into existing simulation code.

\paragraph{\textbf{Running Example: Deterministic Simulations}}
To perform a deterministic simulation of the toy case study we have to remodel it by using only biochemical rewrite rules. This can
be done by phrasing the compartmentalisation by using a different name for the species occurring in the compartment $\textit{IN}$, namely:
$A^{\textit{IN}}$, $B^{\textit{IN}}$ and $C^{\textit{IN}}$. Then, the three non-biochemical rewrite rules $(N_1)$, $(N_2)$ and $(N_3)$ can be converted
into the following biochemical rewrite rules:
$$
\begin{scriptsize}
\begin{array}{@{\!}ccc}
(B_{10}) \;  \TOP :
 A^{IN} \srewrites{d_A} A  %\label{detAexport}
 & (B_{11}) \;
  \TOP :
 B^{IN}\srewrites{d_B}  B %\label{detBexport}
 & (B_{12}) \; \TOP :
  C \srewrites{- d_C} C^{IN} %\label{detCimp}
\end{array}
\end{scriptsize}
$$
The conversion  of the biochemical rules $(B_1),\ldots,(B_9)$ is straightforward. For instance rule $(B_1)$ is
 converted into the two rules:
 $$
\begin{scriptsize}
\begin{array}{c@{\qquad\qquad}c}
(B'_1) \;\;   \TOP : A \srewrites{1}  A \conc A
  & (B''_1) \;\;  \TOP :  A^{\textit{IN}} \srewrites{1}  A^{\textit{IN}} \conc A^{\textit{IN}}
\end{array}
\end{scriptsize}
$$
\begin{figure*}%[t]
\centering
%\subfigure[]{
  \includegraphics[width=.49\textwidth]{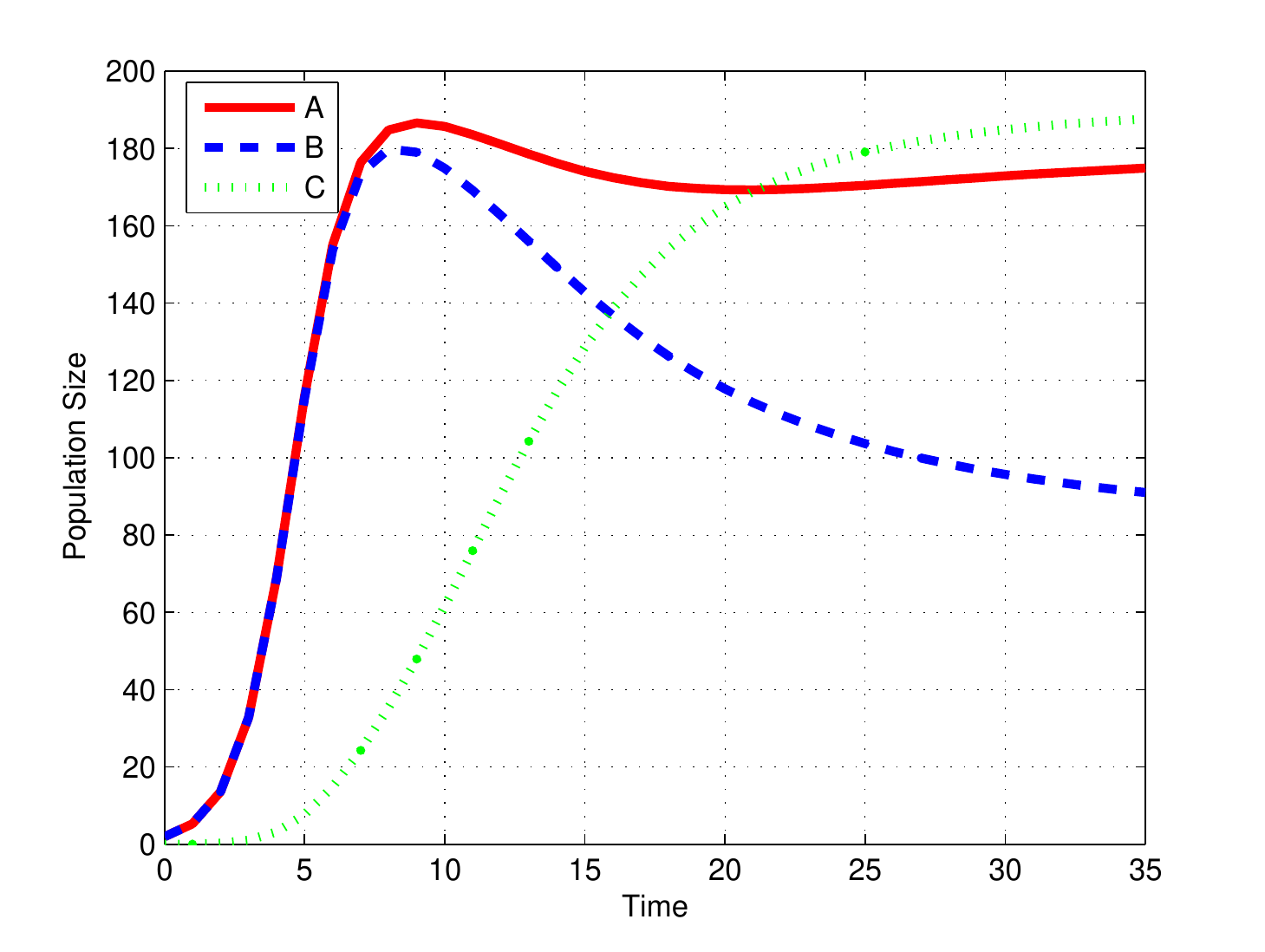}
%  \label{sheepODEIN}}
%  \subfigure[Outside the compartment $IN$]{
  \includegraphics[width=.49\textwidth]{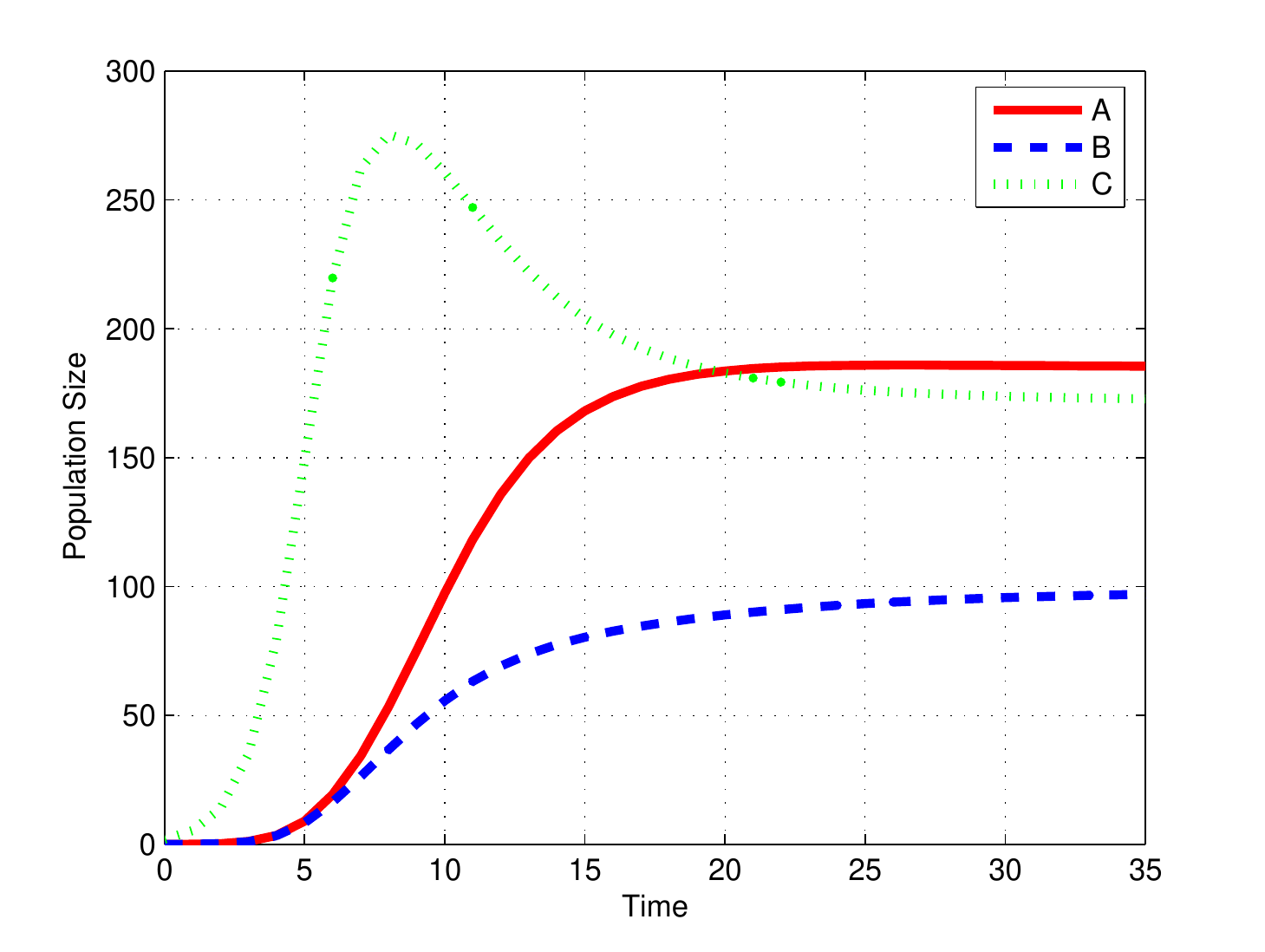}
 % \label{sheepODETopLevel}}
  \caption{Deterministic simulation of the dynamics of the competitive species inside the compartment \textit{IN} (left figure) and outside the compartment (right figure)}
\label{fig:ode_toy}
\end{figure*}
The converted starting term is $2 \times C \conc  2 \times A^{\textit{IN}}\conc 2 \times B^{\textit{IN}}.$
The results of the deterministic simulation are shown in Figure \ref{fig:ode_toy}.

\begin{remark}
The conversion of the original \short\ system for the toy case studies into a \short\ system using only the top level compartment and biochemical
rewriting rules has been straightforward since the original system has a fixed compartment structure. Such a conversion,
as well as the direct representation of the
modelled biological system in terms of ODEs, might be quite complicate (or even impossible) in case of biological systems that during
their evolution may
change the structure of the compartments, maybe by creating a possibly unbounded number of new compartments.
\end{remark}

%%%%%%%%%%%%%%%%%%%%%%%%%%%%%%%%%%%%%%%%%%%%%%%%%%%%%%%%%%%%%

\FORGET{labelled variables resulting in a system of competitive species with the state space defined by:

\begin{equation}
x=(A^{IN},B^{IN},C^{IN},A,B,C)\in R^6_+, x\geq 0
\end{equation}

which represent the $3$ species in the environmental compartment $IN$ and outside it.

The set of non-biochemical rewrite rules $\NN$ was converted into the following set of biochemical rewrite rules $\BB$:

\begin{gather}
A^{IN} \srewrites{d_A} A \label{detAexport}\\
B^{IN}\srewrites{d_B}  B \label{detBexport}\\
C \srewrites{- d_C} C^{IN} \label{detCimp}
\end{gather}

The test case was then modelled as described previously in the paragraph by the following system of ODEs:

\begin{equation}
\begin{array}{lcr}
\frac{dA^{IN}}{dt}=A^{IN} (1-\frac{\mu_{A,A}A^{IN}-\mu_{A,B}B^{IN}-\mu_{A,C}C^{IN}}{K_A})-d_A A^{IN}\\
\frac{dB^{IN}}{dt}=B^{IN} (1-\frac{\mu_{B,A}A^{IN}-\mu_{B,B}B^{IN}-\mu_{B,C}C^{IN}}{K_B})-d_B B^{IN}\\
\frac{dC^{IN}}{dt}=C^{IN} (1-\frac{\mu_{C,A}A^{IN}-\mu_{C,B}B^{IN}-\mu_{C,C}C^{IN}}{K_C})+d_C C^{\intercal}\\
\frac{dA^{\intercal}}{dt}=A^{\intercal} (1-\frac{\mu_{A,A}A^{\intercal}-\mu_{A,B}B^{\intercal}-\mu_{A,C}C^{\intercal}}{K_A})+d_A A^{IN}\\
\frac{dB^{\intercal}}{dt}=B^{\intercal} (1-\frac{\mu_{B,A}A^{\intercal}-\mu_{B,B}B^{\intercal}-\mu_{B,C}C^{\intercal}}{K_B})+d_B B^{IN}\\
\frac{dC^{\intercal}}{dt}=C^{\intercal} (1-\frac{\mu_{C,A}A^{\intercal}-\mu_{C,B}B^{\intercal}-\mu_{C,C}C^{\intercal}}{K_C})-d_C C^{\intercal}
\end{array}
\end{equation}

 }%endFORGET%%%%%%%%%%%%%%%%%%%%%%%%%%%%%%%%%%%%%%%%%%%%%%%%%%%%%%%%

\section{Hybrid Evolution}\label{sect:Hybrid}

The stochastic approach is based on a probabilistic simulation method that
 manages the evolution of exact integer quantities and often requires a huge
computational time to complete a simulation. The ODEs numerical approach computes a unique deterministic and
 %evolution. It expresses a
 fractional evolution of
the species involved in the system and achieves very efficient computations.
%Both techniques, as remarked above, present great advantages with respect to
%the other but also give rise to serious drawbacks.
%
In this section we combine both methods within \short, defining a hybrid simulation technique.

%\subsection{The Hybrid Reduction Semantics}

Given a \short\ system $\QQ$ we partition it into a set of biochemical rewrite rules $\BB$ and a set of non-biochemical rewrite rules $\NN$.  Rules in $\NN$
are always applied by using the stochastic method. Rules in $\BB$ might be applied with the ODEs approach. In general $\BB$ might contain both rules that
model evolution of large numbers of molecules according to very fast reactions (whose execution is suitable to be correctly computed with ODE) and rules that model very slow reactions or reactions that involve a very
small number of reagents. In the latter case 
%case of reactions that are very slow or involve a very small number of reagents 
it is convenient to compute the execution of
the associated rule according to the stochastic approach. 

According to the state of the system, a rule might be dynamically interpreted either as stochastic or deterministic. For instance, during a simulation, it might happen that a given biochemical rewrite rule $\ell:
\bar{a}_i \srewrites{k_i} \bar{b}_i \in \BB$ is applied initially according to the stochastic semantics, since the associated compartment contains a very
small number of reagents. After the system has evolved for some time, however, the concentration of the reagents involved in the rule can be substantially
increased and it becomes convenient to model the corresponding reaction according to the deterministic approach.

 Actually, at the beginning of each simulation
step we build, for each compartment in the term, a system of ODEs for the simulation of the biochemical rules in that compartment which (1) are
sufficiently fast and (2) involve reagents with a sufficient concentration. For the remaining rules the evolution is determined by the stochastic simulation
algorithm.

In order to describe the hybrid semantics we assume that, given a \short\ term $\ov{t}$, each compartment of $\ov{t}$ is univocally identified by an
index $\iota$. The index of the (implicit) compartment at the top level will be denoted by $\iota_0$. The \emph{biochemical reagents} of a compartment $(\ov{a}\into \ov{t})^\ell$ with index
$\iota$, written $\BR(\iota)$, are expressed by the multiset of the atomic elements appearing in the top level of $\ov{t}$. For example, given the term
\[
\ov{t} = a \conc a \conc b \conc (c \into (d \conc e \into \emptyseq)^{\ell'} \conc f )^\ell \conc (c \into \conc f \conc g )^\ell
\]
and assuming that the compartment $(c \into (d \conc e \into \emptyseq)^{\ell'} \conc f )^\ell$ has index $\iota_1$, the compartment $(d \conc e \into
\emptyseq)^{\ell'}$ has index $\iota_2$ and the compartment $\conc (c \into \conc f \conc g )^\ell$ has index $\iota_3$, we have that $\BR(\iota_0)=a \conc a
\conc b$, $\BR(\iota_1)=f$, $\BR(\iota_2)=\emptyseq$ and $\BR(\iota_3)=f \conc g$.

\begin{figure}[t]
\hrule $\;$ \\
Let $\ov{t}$ denote the whole term and let $I$ denote the set of compartment indexes occurring in $\ov{t}$.
\begin{enumerate}
\item For each compartment $\iota\in I$:   %in  $\ov{t}$
\begin{itemize}
\item Let $\ell$ be the label of $\iota$, let $\DD_\iota=\BB_\ell$ and let $\RR_\iota=\emptyset$.
\item For each biochemical rule %$B_i=(\ell,\bar{a}_i,k_i, \bar{b}_i) \in \DD_\iota$,
$B_i=\ell:\bar{a}_i \srewrites{k_i} \bar{b}_i \in \DD_\iota$
let $\bar{a}_i = a_{i_1}\;\ldots\;a_{i_{r_i}}$ ($r_i\ge 1$) and let $[a_{i_1}]^\iota,\ldots,[a_{i_{r_i}}]^\iota$ denote the concentrations of the species
occurring in $\bar{a}_i$ within the multiset $\BR(\iota)$. Let $K_i^{\iota}$ be the rate of the application of rule $B_i$ in $\iota$. Namely,
$K_i^{\iota}=k_i \cdot [a_{i_1}]^\iota\cdot \ldots \cdot[a_{i_{r_i}}]^\iota$. If $K_i^{\iota} < \phi$ or $\textit{min}\{[a_{i_1}]^\iota,\ldots,
[a_{i_{r_i}}]^\iota\} < \psi$ remove $B_i$ from $\DD_\iota$ and put it into $\RR_\iota$.
\end{itemize}
\item Considering the rules in $\bigcup_{\iota \in I} \RR_\iota \cup \NN$ select according to Gillespie's method and to the semantics given in Section~\ref{SECT:STO_SEM}
a stochastic transition step $C[P\sigma] \ltrans{k \cdot \OO(R, C[P\sigma],C[O\sigma])} C[O\sigma])$, where $R=\ell:P \srewrites{k}  O
\in \RR_{\iota'} \cup \NN_\ell$ 
%is the next rule to be applied (for some compartment $\iota'$ of type $\ell$). 
\\
Let $\tau$ be the corresponding time interval.$^\dagger$
\item
For each compartment $\iota$ in $I$:
\begin{itemize}
\item
Let $\Ode_{\iota}$ denote the system of ODEs for the rules in $\DD_\iota$ in the compartment $\iota$ as explained in Section~\ref{SECT:ODE_SEM} without
considering, in the compartment $\iota'$ where the stochastic transition step takes place, the active reagents appearing in the left part $P$ of the
stochastically applied rule. (If $\DD_\iota=\emptyset$ then $\Ode_{\iota}=\emptyset$.)
\item Apply the system of ODEs $\Ode_{\iota}$ to the biochemical reagents $\BR(\iota)$
 of the compartment for a time duration $\tau$.
\end{itemize}
\item Update the term $\ov{t}$ according to the right part $O$ of the chosen stochastic rule and to the applications of the systems of ODEs.
\end{enumerate}
\underline{\hspace{4cm}}\\
 \footnotesize{$^\dagger$ In some rare case, it may happen that no rule in $(\bigcup_{\iota \in I} \RR_\iota)\cup\NN$ is applicable.
In such cases the evolution of the system must be determined for some time $\tau$ according to the deterministic semantics only. In our implementation we
choose as $\tau$ the maximum time calculated by Gillespie's algorithm for each of the applicable biochemical rules in $\bigcup_{\iota \in I}\DD_\iota$.}
 \caption{Steps performed by an hybrid simulation iteration}
\label{fig:CWM-hybrid-step}
\end{figure}

A basic point of our hybrid approach is the criterium to determine, at each computation stage, the reductions to compute in stochastic or in a deterministic way. In this paper we have chosen simply to put a threshold on the number of possible reagents and on the speed of the reaction, but other more sophisticated criteria should be investigated.

Given a term $\ov{t}$ to reduce, a rate thresholds $\phi$ and a concentration threshold $\psi$, each iteration of the
hybrid reduction semantics performs the four steps listed in Figure~\ref{fig:CWM-hybrid-step}, where,
%the set of biochemical rules is
%$\BB=\{B_1,\ldots,B_n\}$ $(n\ge 1)$ and,
for every label $\ell$, the subsets of $\BB$ and $\NN$ containing the rules with label $\ell$ are denoted by
$\BB_\ell$ and $\NN_\ell$, respectively. The first step identifies, for each compartment $\iota\in I$ (where $I$ is the set of all compartment indexes occurring in $\ov{t}$), two disjoint sets of biochemical rules, namely
$\DD_\iota$ (to be applied deterministically) and $\RR_\iota$ (to be applied, together with the rules in $\NN$, according to the stochastic method). The
second step selects, considering only the rules in $\bigcup_{\iota \in I} \RR_\iota \cup \NN$ the next rule to be applied stochastically. The third step computes a system of ODEs $\Ode_\iota$ for each compartment $\iota\in I$ and applies
the ODEs for the time duration selected by the stochastic step. The fourth step updates the terms according to the results of the simulation.\footnote{Note that since ODEs deal with fractional quantities, a rounding operation will be needed before computing the next stochastic step.}

In general, if reactions are fast enough, the deterministic ODEs simulation approximate better the exact stochastic simulations. This is the idea behind the use of the threshold $\phi$. The use of $\psi$, instead, allows to prevent the rounding approximation error that may derive when we are dealing with species at low concentrations. Combined together, the thresholds $\phi$ and $\psi$ affect the level of approximation we want to use in our simulations. Notice that with $\phi=+\infty$ all reactions will be considered \emph{too slow} and the simulation will be computed with the purely stochastic method.

\paragraph{\textbf{Running Example: Hybrid Simulations}}
The hybrid simulations were performed by using thresholds $\phi = 60$ and $\psi = 60$, which were determined as feasible thresholds to catch the initial stochastic effects
for the multistable behaviour of the dynamic system.

\begin{figure*}[t]
\centering
%\subfigure[First experiment]{
  \includegraphics[width=.45\textwidth]{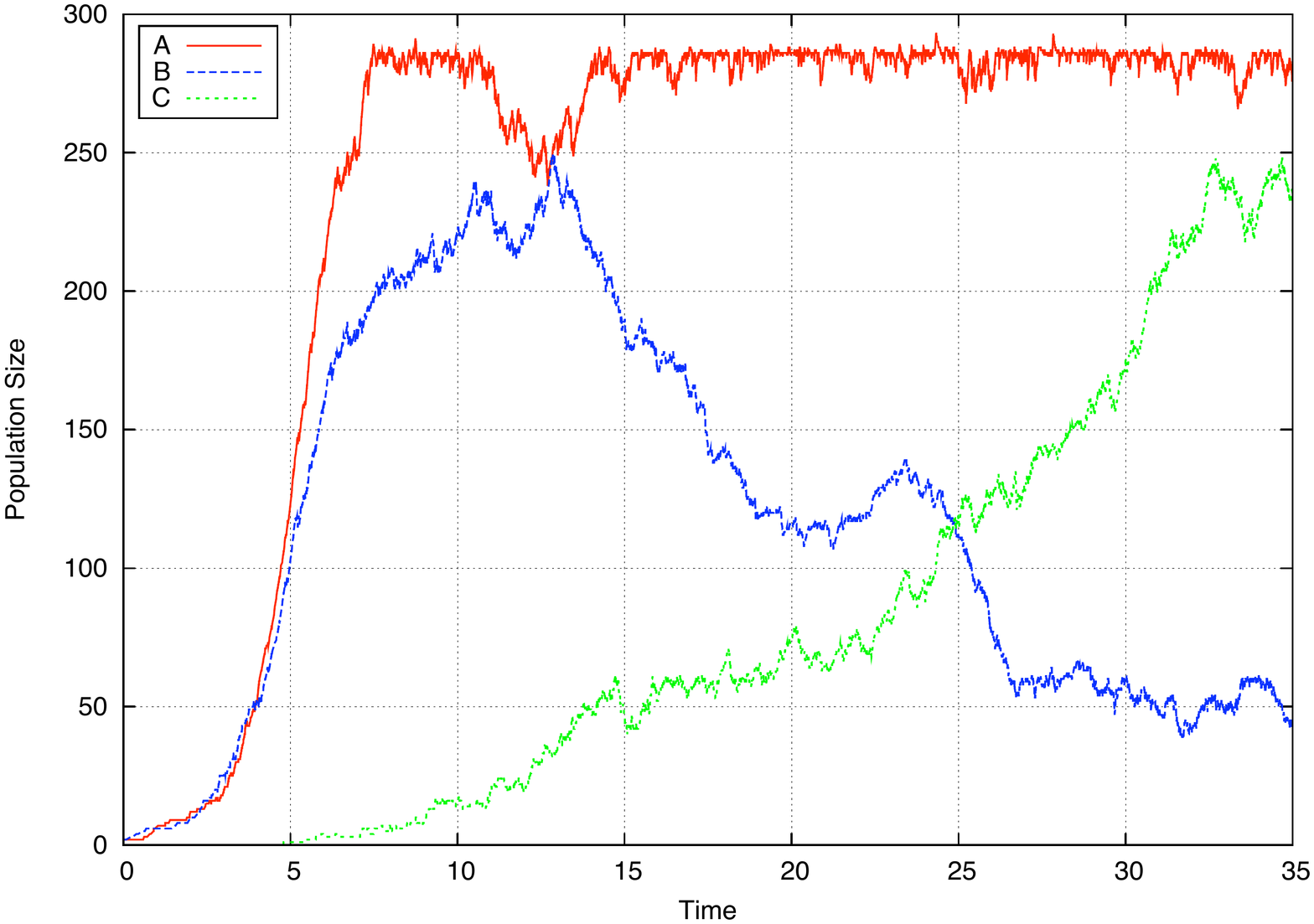}
%  }\centering
%\subfigure[In $\TOP$]{
  \includegraphics[width=.45\textwidth]{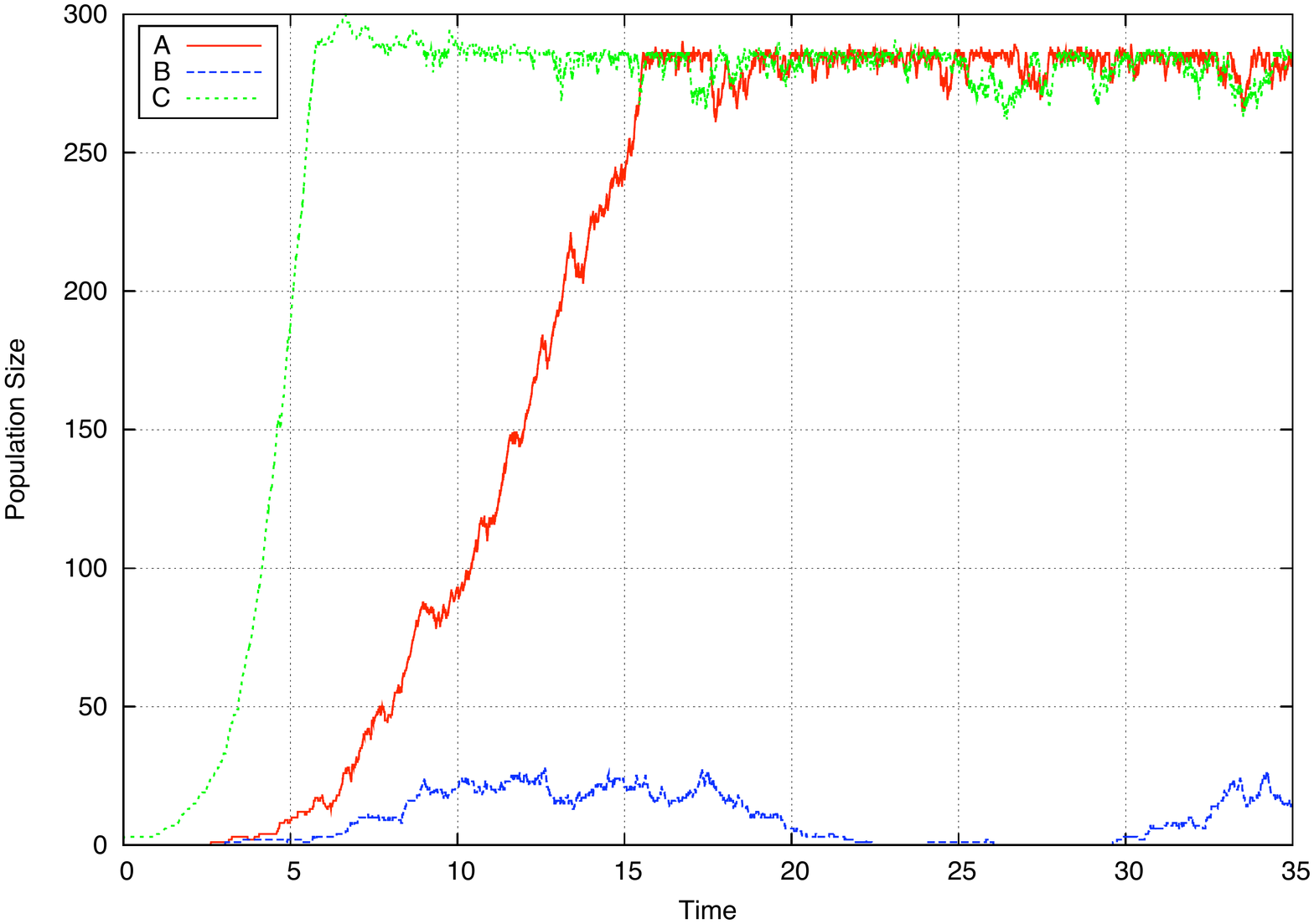}
  %}
  \hrule
 %\centering
%\subfigure[Second experiment]{
  \includegraphics[width=.45\textwidth]{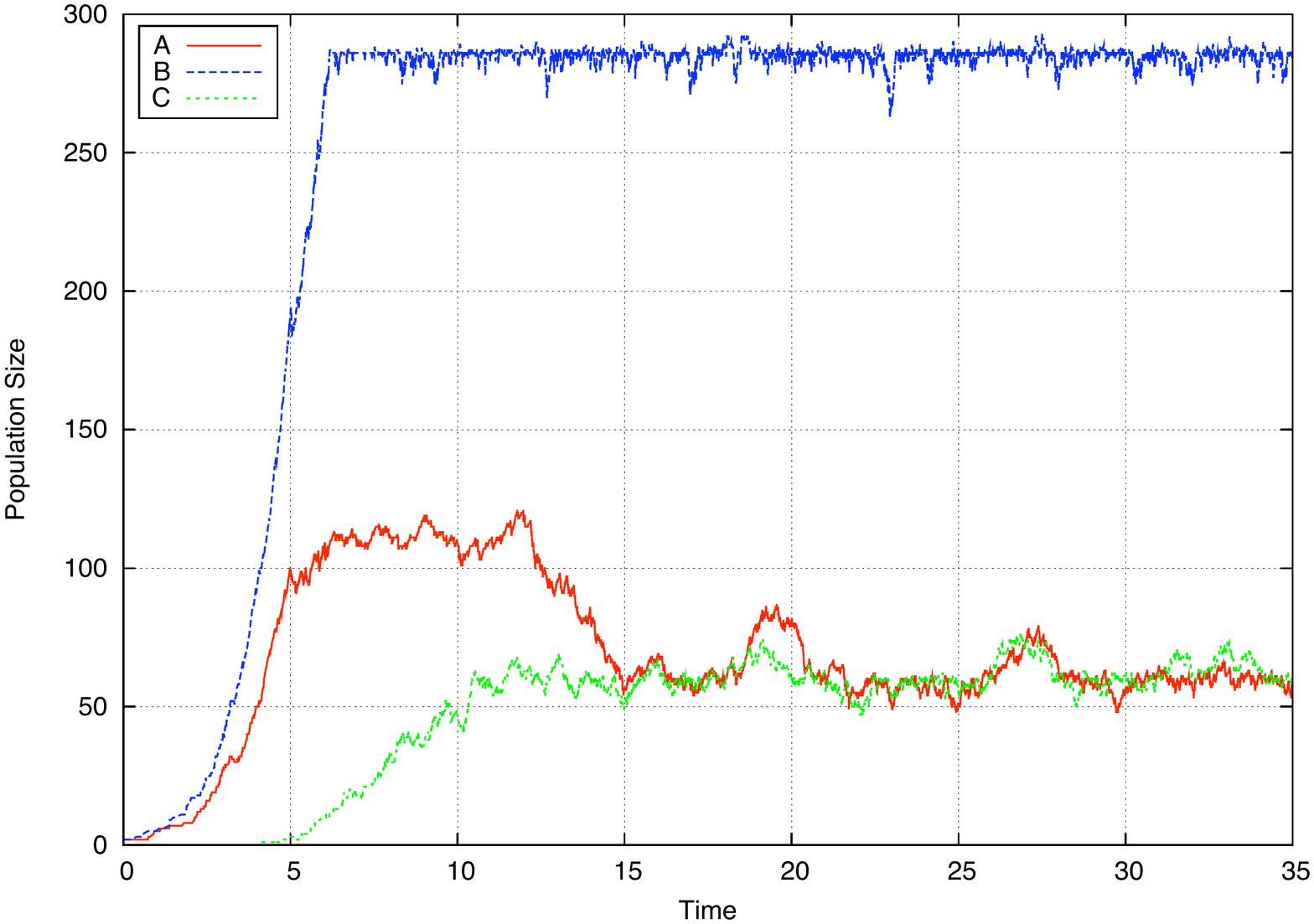}
  %}\centering
%\subfigure[In $\TOP$]{
  \includegraphics[width=.45\textwidth]{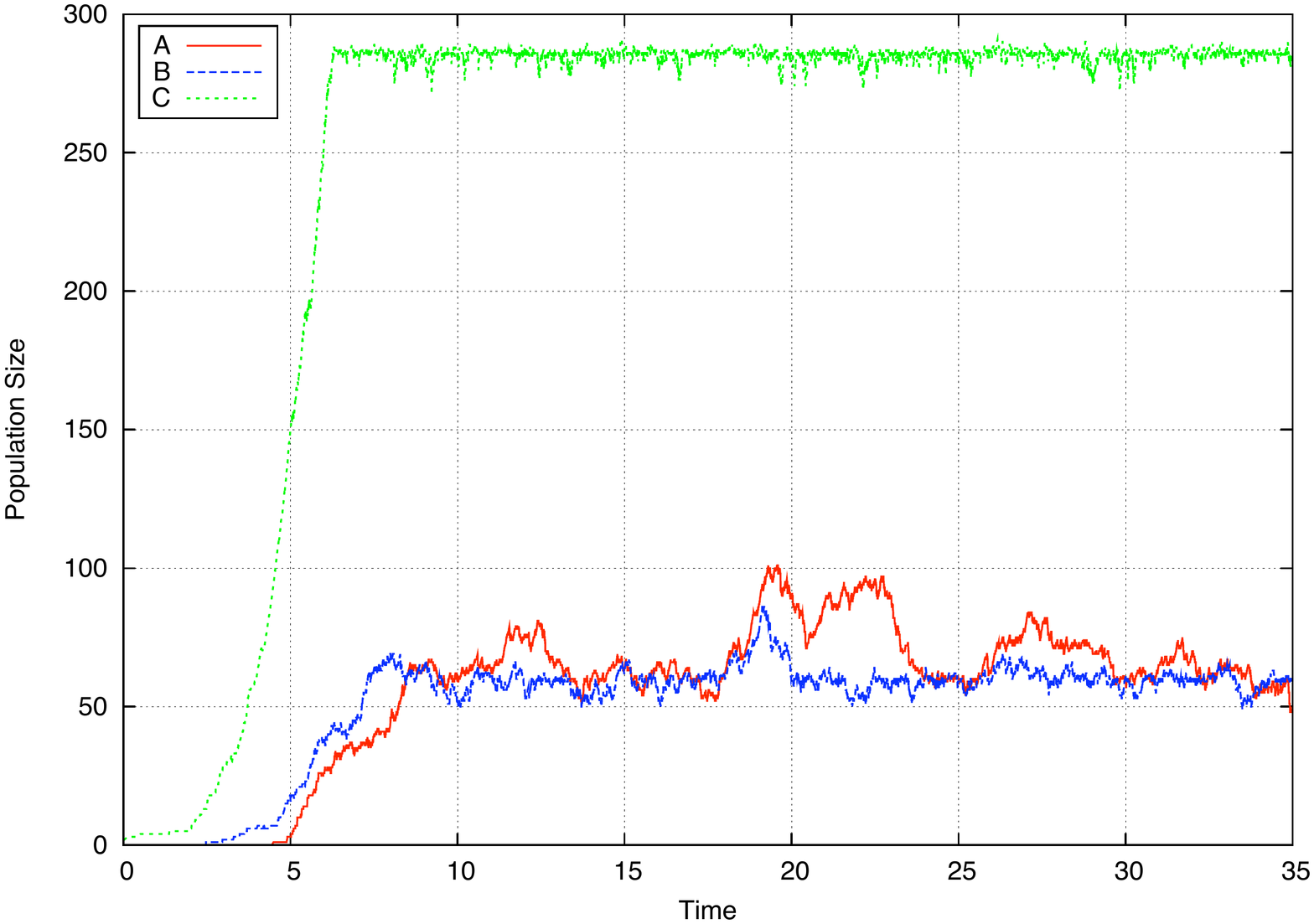}
  %}
\caption{Two different runs of the hybrid simulations showing the different behaviour of the dynamics of the competitive species
inside the compartment \textit{IN} (on the left side) and outside the compartment (on the right side)}
\label{fig:hybr_toy}
\end{figure*}

In Figure~\ref{fig:hybr_toy} we report two runs of the hybrid simulations showing two different evolutions of the competitive species inside the
compartment \textit{IN} (on the left side) and outside the  compartment (on the right side).
The hybrid method allows the synthesis of these evolutions of
the system (among the many possible ones) by identifying, through the choice of the thresholds $\phi $ and $\psi$, the system state
in which the populations are considered large enough to allow differential equations to take the government of the system evolution.  In this example in fact, as it is easy to verify,  while the destiny of the species is determined by the stochastic bootstrap, after the
populations have reached the chosen threshold their evolution becomes regulated only by ODE transitions. The deterministic computation of the evolution of
the system then tracks an average
behaviour reducing the computational load. At this point, in fact, stochastic fluctuations, can be considered irrelevant for the description of the overall behaviour of the system.  The moment in which the simulations move from the stochastic method to the deterministic one can be easily captured in the graphs shown in Figure \ref{fig:hybr_toy}.

%the hybrid approach introduces aspects of randomness due to the interactions among the species,
%especially in the early stages of the simulation, when few initial individuals of the system
%can interact departing from a statistical average behaviour. then, when the species reach
%statistically large volumes, system fluctuations are neglected in order to quickly find
%the steady state of the system.

The hybrid method provides a faster final solution with respect to the pure stochastic approach
by a factor of $10$ giving a very satisfiable qualitative analysis of its behaviour.

\section{A Real Model of Different Cellular Fate}\label{sect:tat}

To assess the soundness and efficiency of our hybrid approach on a real biological
problem we decided to apply it to a well known system where stochastic effects
play a fundamental role in determining its development: the HIV-1 transactivation mechanism.

After a cell has been infected, the retrotransposed DNA of the virus is integrated in the host genome and it begins its transcription in \textit{mRNA} and
then the translation to yield viral proteins; the initial speed of this mechanism, however, is fairly slow. The speedup of the viral production process is
determined by a regulation system driven by the viral protein \textit{TAT}: this protein is capable of binding cellular factors of the host to produce the
\textit{pTEFb} complex which in its acetylated form is able to bind to the integrated viral genome and speed up the transcription machinery, thus ending
in more viral proteins and, therefore, more \textit{TAT}, determining a positive loop.

The time scale during which this loop is triggered is affected by many factors: the initial low \textit{TAT} production and the rate of its degradation,
the equilibrium between the active (acetylated) and inactive form of \textit{pTEFb} and so on. As a consequence, the stochastic oscillations in this
events are considered pivotal in determining when viral proteins are produced in a sufficient quantity to determine cellular lysis and viral spreading.
Since HIV is known to stay dormant and inactive in some types of cells and since the time between the infection and the high viral production rate related to the
active phase of AIDS is variable, this transactivation mechanism is of great interest.

We decided to follow the direction taken in a previous study about this system (see~\cite{WBTAS05}), in which an experimental setting is developed where a
fluorescent protein, \textit{GFP}, is the only one encoded by an engineered viral genome, along with TAT. In~\cite{WBTAS05} they were able to identify
different evolutions in the \textit{GFP} level over time: cellular clones with exactly the same genome showed two different behaviour, one produced a high
quantity of \textit{GFP} (they called it ``bright'') and the other one with very little \textit{GFP} (``off''). This work also reported that a purely
stochastic simulation was able to individuate this bifurcation; a later work (see~\cite{GCPS06}) confirmed these results performing purely stochastic and
mixed deterministic-stochastic simulations.

\begin{figure}[t]
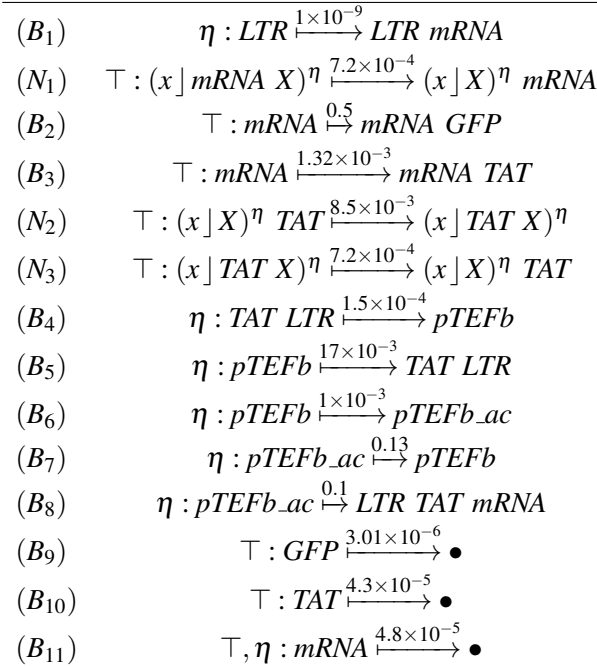

\hrule
$
\begin{array}{lc}
(B_1) & \eta : \textit{LTR} \xsrewrites{1 \times 10^{-9}} \textit{LTR} \conc \textit{mRNA} %\label{basal}
\\
(N_1) & \TOP : (x \into \textit{mRNA} \conc X)^\eta \xsrewrites{7.2 \times 10^{-4}} (x \into X)^\eta \conc \textit{mRNA} %\label{RNAexport}
\\
(B_2) & \TOP : \textit{mRNA} \xsrewrites{0.5} \textit{mRNA} \conc \textit{GFP} %\label{prodGFP}
\\
(B_3) & \TOP : \textit{mRNA} \xsrewrites{1.32 \times 10^{-3}} \textit{mRNA} \conc \textit{TAT} %\label{prodTAT}
\\
(N_2) & \TOP : (x \into X)^\eta \conc \textit{TAT} \xsrewrites{8.5 \times 10^{-3}} (x \into \textit{TAT} \conc X)^\eta %\label{TATimp}
\\
(N_3) & \TOP : (x \into \textit{TAT} \conc X)^\eta \xsrewrites{7.2 \times 10^{-4}} (x \into X)^\eta \conc \textit{TAT} %\label{TATexp}
\\
(B_4) & \eta : \textit{TAT} \conc \textit{LTR} \xsrewrites{1.5 \times 10^{-4}} \textit{pTEFb} %\label{trans}
\\
(B_5) & \eta : \textit{pTEFb} \xsrewrites{17 \times 10^{-3}}  \textit{TAT} \conc \textit{LTR}%\label{rev_trans}
\\
(B_6) & \eta : \textit{pTEFb} \xsrewrites{1 \times 10^{-3}} \textit{pTEFb\_ac} %\label{ac1}
\\
(B_7) & \eta : \textit{pTEFb\_ac} \xsrewrites{0.13} \textit{pTEFb} %\label{ac2}
\\
(B_8) & \eta : \textit{pTEFb\_ac} \xsrewrites{0.1} \textit{LTR} \conc \textit{TAT} \conc \textit{mRNA} %\label{fast}
\\
(B_9) & \TOP : \textit{GFP} \xsrewrites{3.01 \times 10^{-6}} \emptyseq %\label{degrGFP}
\\
(B_{10}) & \TOP : \textit{TAT} \xsrewrites{4.3 \times 10^{-5}} \emptyseq %\label{degrTAT}
\\
(B_{11}) & \TOP, \eta: \textit{mRNA} \xsrewrites{4.8 \times 10^{-5}} \emptyseq %\label{degrmRNA}
\end{array}
$
 \caption{ \CalculusShortName\ rules for the TAT transactivation system}
\label{fig:TAT-rules}
\end{figure}

Since \short\ systems are able to represent compartments, we slightly modified the original set of rules used in these works to explicitly represent the
cytoplasm and the nucleus of an infected cell; all the kinetic rates were maintained, the one for \textit{TAT} nuclear import has been determined from
the literature (see~\cite{NLRB09}). The set of rules we adopted is given in Figure~\ref{fig:TAT-rules}, where we refer to the cytoplasm as the $\TOP$
compartment while $\eta$ is the label used for the nucleus. As regards the rules:
 $(B_1)$ %\ref{basal}
 represents the slow basal rate of viral \textit{mRNA} transcription;
 $(N_1)$ %\ref{RNAexport}
 describes the \textit{mRNA} export from the nucleus to the cytoplasm;
 $(B_2)$ %\ref{prodGFP}
 and $(B_3)$ %\ref{prodTAT}
 express the translations of this \textit{mRNA} into
\textit{GFP}  and \textit{TAT} proteins, respectively;
 $(N_2)$ %\ref{TATimp}
 and $(N_3)$ %\ref{TATexp}
 represent the nuclear import and export of \textit{TAT};
 $(B_4)$ and $(B_5)$%\ref{trans}
 models
the binding and unbinding of \textit{TAT} with (not represented here) host cellular factors and the viral genome portion \textit{LTR} that forms \textit{pTEFb} which, when
acetylated (by rule $(B_6)$) %\ref{ac1}
 determines an higher transcriptional activity, which is represented in $(B_8)$ %\ref{fast}
 by the unbinding that releases
 \textit{LTR} and \textit{TAT} and creates an \textit{mRNA} molecule (note the higher rate with respect to
   $(B_1)$); %\ref{basal}
   $(B_7)$  %\ref{ac2}
   represents the \textit{pTEFb} deacetylation and
 $(B_9)$, %\ref{degrGFP}
 $(B_{10})$ %\ref{degrTAT}
 and $(B_{11})$ %\ref{degrmRNA}
 model the degradation processes of the proteins and the \textit{mRNA} (note that \textit{mRNA} degrades both in the nucleus
and in the cytoplasm, the other proteins only degrade in the cytoplasm; also note how the compartment labelling mechanism allows to express this fact in a
simple and elegant way).

\begin{figure*}%[t]
\centering
%\subfigure[Pure stochastic] {
\includegraphics[width=.45\textwidth]{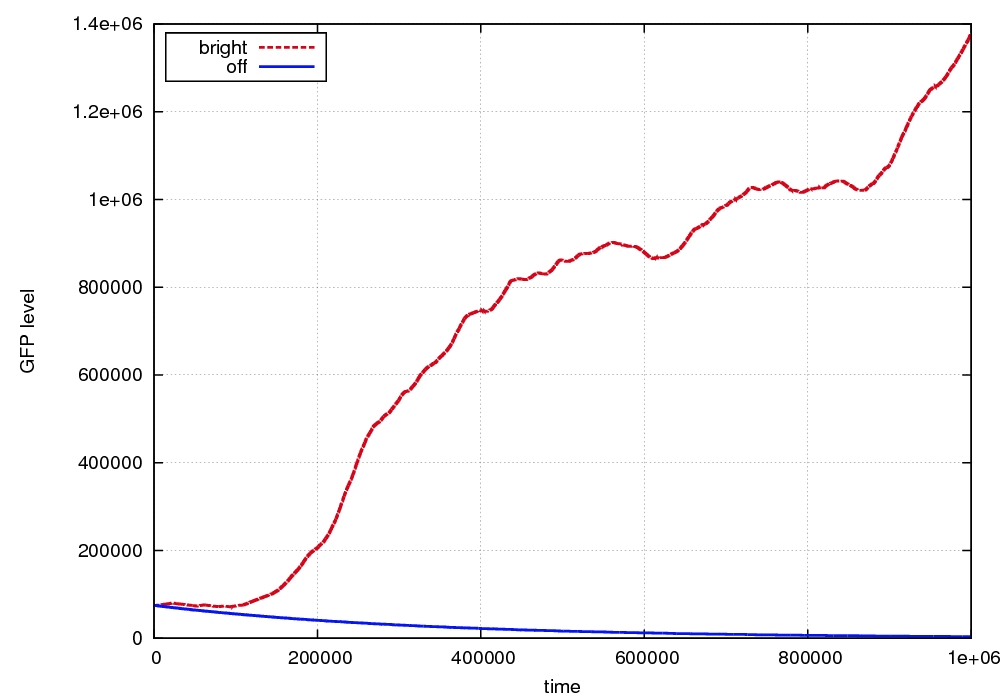}
%}
%\centering
%\subfigure[Hybrid]{
\includegraphics[width=.45\textwidth]{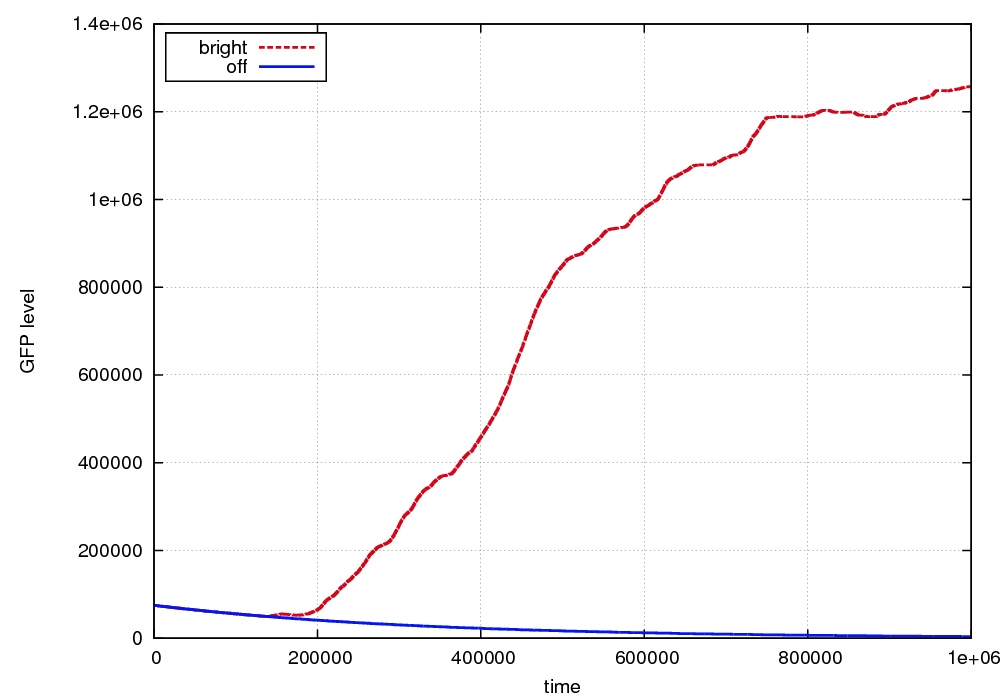}
%}
\caption{Two different simulations pure stochastic (on the left) and hybrid (on the right) started with the same parameters: ``bright'' and ``off'' behaviour}
\label{fig:giano}
\end{figure*}

We performed 100 purely stochastic simulations (i.e. setting $ \phi = +\infty$) and 100 hybrid simulations (using $ \phi = 0.5$ and $\psi = 10$). The values of $\phi$ and $\psi$ for the hybrid simulation were
determined, after a few experiments, as the right values to be able to grasp the stochastic effects for this system. The initial term of our simulations is represented by the
\short\ term
$$
75000 \times \textit{GFP} \conc 5 \times \textit{TAT} \conc (\emptyseq \into \textit{LTR})^\eta ,
$$
 while the time interval of our simulations has been fixed to
$10^6$ seconds (the same parameters are used in~\cite{WBTAS05,GCPS06}). Both our stochastic and hybrid simulations clearly showed the two
possible evolutions of the system which correspond to the ``bright'' and the ``off'' cellular populations (in order to display the double destiny, almost
all the biochemical rewrite rules have to be simulated with the stochastic approach). As could be seen in Figure~\ref{fig:giano}, the hybrid simulations
are comparable to the purely stochastic ones and, even with the relatively high thresholds used in this particular case, the hybrid simulations were
computationally more efficient (almost 40\% faster).\footnote{Comparisons are made using the same stochastic engine, in both cases with no particular optimization.}

\section{Conclusions}\label{conc}

As we have seen, \short\ allows to model cellular interaction, localisation and membrane structures.
Other formalisms were developed to describe membrane
systems. Among them we cite Brane Calculi~\cite{Car05} and
P-Systems~\cite{P02}.

\short\ can describe situations that cannot be easily captured by the
previously mentioned formalisms, which consider membranes as atomic
objects (extensions of P-Systems with objects on membranes can be found in~\cite{BCRRS08,CS08}). Representing the membrane structure as a multiset of the elements
of interest allows the definition of different functionalities
depending on the type and the number of elements on the membrane
itself.

In this paper we have defined a hybrid simulation technique for systems described in CWC, which combines the stochastic approach with the deterministic one obtained through ODEs. The method alternates discrete transitions, computed probabilistically according to the stochastic method, and continuous transitions, computed in a deterministic way by a set of ODEs. Our technique turns out to accurately capture the dynamics of systems that exhibit stochastic effects and takes advantage, whenever the deterministic approach is applicable, of the efficiency of the ODEs simulation method.

The running examples used to make comparisons between the different quantitative simulation methods, and the HIV-1 transactivation mechanism are challenging tests for our hybrid methodology. On the one hand, the running example shows that, from the methodological point of view, several runs of hybrid simulations on a model of dynamic competitive populations allow to ``synthesize'' a set of stochastic experiments avoiding the statistical assumptions about the initial distributions of the parameters that are needed in a purely deterministic or purely stochastic analysis. On the other hand, the simulation of the HIV-1 transactivation mechanism follows a simulation which is \emph{almost} purely stochastic: only a few rules pass the threshold condition, thus the computational gain of the deterministic approach is, in this particular case, very limited (even if still sensible).

%Actually, we might have chosen a better example for which Hybrid \short\ does a perfect work. Just consider a simple %\short\ term:

%$$\ov{t}=100000 \times a \conc  1000 \times e \conc (r \into 100000 \times b \conc 1000 \times e)^{\ell}$$

%with rules:

%$$
%\begin{array}{lc}
%(B_1) & \TOP : a \conc e \xsrewrites{0.1} a' \\
%(B_2) & \TOP : a'  \xsrewrites{0.01} a \conc e \\
%(B_3) & \ell : b \conc e \xsrewrites{0.3} b' \\
%(B_4) & \ell : b'  \xsrewrites{0.03} b \conc e\\
%(N_1) & \TOP : (r \into b' \conc X)^\ell \xsrewrites{1 \times 10^{-4}} b' \conc (r \into  X)^\ell\\
%(B_5) & \TOP : b' \conc e' \xsrewrites{0.1} c \\
%\end{array}
%$$

%where substrates $a$ and $b$ may bind with enzyme $e$ either at the top level or within a cellular compartment creating, %respectively, the $a'$ and $b'$ complexes. A transporter $r$ on the membrane pumps the complexes $b'$ outside the cell, %where they can interact with $a'$ generating $c$. This example is perfectly suited for Hybrid CWC: we have a slow %stochastic rule moving elements between compartments and a few, very fast, chemical reaction. Actually, the results of %the purely stochastic and the hybrid simulations do coincide (see Figure~\ref{???}). However, the computational cost of %the stochastic simulation (s-seconds) is k-times more expensive than the the hybrid simulation (h-seconds).

Compartment labels introduced in this paper are a novelty with respect to the original \short\ calculus presented in~\cite{preQAPL2010}. As we have seen, these labels are necessary when building a system of ODEs for a compartment of type $\ell$. However, we might exploit these labels as an intrinsical information about the properties of a compartment. For example, assuming that compartments of the same type have approximatively the same volume, we might use the compartment type to define a set of biochemical rules whose kinetics incorporate the information about the volume of the compartment on which the rule could be applied. Suppose, in practice, to analyse a system in which two different kind of cells may interact. Let's call $\ell_1$ and $\ell_2$ the compartment types of the two kinds of cells. Suppose, then, that particles $a$ and $b$ are free to float between these cells and the top level interspace hosting all the cells. Finally, particles $a$ and $b$ may interact by complexation and produce the particle $c$. If it holds that the top level interspace on which the different cells float has around 100x the volume of a cell of type $\ell_1$ and if a cell of type $\ell_1$ has around 3x the volume of a cell of type $\ell_2$, we can express the different speeds of the $a-b$ complexation in the different compartments (according to their volumes) with the three following rules:
$$\TOP:a \conc b \srewrites{k} c, \qquad \ell_1:a \conc b \srewrites{k\cdot100} c, \qquad \ell_2:a \conc b \srewrites{k\cdot 300} c.$$

Actually, it is crucial to consider in detail the volumes of the involved compartments and to consider adequate kinetics for the biochemical rules
used to simulate the system behaviour. We notice, in particular, that the approach based on ODEs directly translates chemical reactions
into mathematical equations and computes the concentrations over time of the involved species (usually the \emph{molar concentration},
which denotes the number of moles of a given substance per liter).
Models based on the stochastic approach, instead, simulate the activity of each single individual involved in the evolution of the system.
Such a delicate difference between the two methods should be carefully taken into account when developing the set of rules to be simulated
with the hybrid approach. The version of \short\ with labelled compartments presented in this paper simplifies this kind of analysis and allows for more accurate simulations.
%Relating to this fact, the application presented in Section~\ref{sect:tat} uses the kinetics presented %in~\cite{WBTAS05,NLRB09}, which were already adapted according to the approximate volume of an infected cell %(\textbf{Elena mi correggi o e' giusto?}).

%We are also considering to enrich our analysis framework by taking into account statistical
%model checking as done in~\cite{BCM09} for stochastic CLS models, and with techniques to approximate the value of %unknown kinetics in the stile of~\cite{LMT04,LMT07,CGL09}.

\subsection*{Acknowledgements}\label{ack}

We gratefully acknowledge the helpful comments and suggestions received from the anonymous reviewers of MeCBIC 2010.

The authors also wish to thank Sergio Rabellino and the ICT staff of the Computer Science Department of the University of Turin
for providing technical support and assistance in running the simulations. Finally, we thank Prof. Nello Balossino (University of Turin) who made us available the computing resources of the laboratory Segnali e Immagini ``G. Tamburelli''.

\bibliographystyle{eptcs}

\bibliography{fmb}

\end{document}